\documentclass[superscriptaddress,showkeys,preprint]{revtex4-1}
\usepackage{graphicx}
\usepackage{amssymb,amsfonts,amsmath}

\renewcommand{\vec}[1]{{\mathbf{#1}}}
\newcommand{\va}{\vec a}
\newcommand{\vk}{\vec k}
\renewcommand{\vr}{\vec r}

\newcommand{\del}{\partial}
\newcommand{\eps}{\epsilon}

\newcommand{\rhoTail}{\rho_{\text{tail}}}
\newcommand{\sigmaTail}{\sigma_{\text{tail}}}
\newcommand{\epsTail}{\eps_{\text{tail}}}

\newcommand{\muHead}{\mu_{\text{head}}}
\newcommand{\sigmaHead}{\sigma_{\text{head}}}
\newcommand{\epsHead}{\eps_{\text{head}}}

\newcommand{\kB}{k_{\text{B}}}
\newcommand{\muex}{\mu^{\text{ex}}}

\newcommand{\hTwiddle}{{\tilde h}}

\newcommand{\xiTwiddle}{{\tilde\xi}}
\newcommand{\hDot}{{\dot h}}
\newcommand{\hDotTwiddle}{{\dot\hTwiddle}}
\newcommand{\hDDotTwiddle}{{\ddot\hTwiddle}}

\newcommand{\HTwiddle}{{\tilde H}}
\newcommand{\HpTwiddle}{{\tilde H'}}

\begin{document}

\title{Extended surfaces modulate and can catalyze hydrophobic effects}

\author{Amish J. Patel}
\affiliation{Howard P. Isermann Department of Chemical \& Biological Engineering, and Center for Biotechnology and Interdisciplinary Studies, Rensselaer Polytechnic Institute, Troy, NY 12180, USA}

\author{Patrick Varilly}
\affiliation{Department of Chemistry, University of California, Berkeley, CA 94720, USA}

\author{Sumanth N. Jamadagni}
\author{Hari Acharya}
\affiliation{Howard P. Isermann Department of Chemical \& Biological Engineering, and Center for Biotechnology and Interdisciplinary Studies, Rensselaer Polytechnic Institute, Troy, NY 12180, USA}

\author{Shekhar Garde}
\altaffiliation{To whom correspondence should be addressed. Email: gardes@rpi.edu or chandler@berkeley.edu} 
\affiliation{Howard P. Isermann Department of Chemical \& Biological Engineering, and Center for Biotechnology and Interdisciplinary Studies, Rensselaer Polytechnic Institute, Troy, NY 12180, USA}

\author{David Chandler}
\altaffiliation{To whom correspondence should be addressed. Email: gardes@rpi.edu or chandler@berkeley.edu} 
\affiliation{Department of Chemistry, University of California, Berkeley, CA 94720, USA}

\begin{abstract}
Interfaces are a most common motif in complex systems.
To understand how the
presence of interfaces affect hydrophobic phenomena, we use molecular
simulations and theory to study hydration of solutes at interfaces.  The solutes
range in size from sub-nanometer to a few nanometers.  The interfaces are
self-assembled monolayers with a range of chemistries, from
hydrophilic to hydrophobic. We show that the driving force for
assembly in the vicinity of a hydrophobic surface is weaker than that
in bulk water, and decreases with increasing temperature, in contrast
to that in the bulk. We explain these distinct features in terms of an
interplay between interfacial fluctuations and excluded volume effects---the physics encoded in Lum-Chandler-Weeks theory [J.~Phys.~Chem.~B
\textbf{103} 4570--4577 (1999)].  Our results suggest a catalytic role for
hydrophobic interfaces in the unfolding of proteins, for example, in
the interior of chaperonins and in amyloid formation.
\end{abstract}


\maketitle

Hydrophobic effects are ubiquitous and often the most
significant forces of self-assembly and stability of nanoscale
structures in liquid matter, from phenomena as simple as micelle
formation to those as complex as protein folding and aggregation
\cite{tanford_book,kauzmann}. These effects depend importantly on
lengthscale \cite{stillinger,LCW,DC_nature05}. Water molecules near
small hydrophobic solutes do not sacrifice hydrogen bonds, but have
fewer ways in which to form them, leading to a large negative entropy
of hydration. In contrast, hydrogen bonds are broken in the hydration
of large solutes, resulting in an enthalpic penalty. The crossover from
one regime to the other occurs at around $1\,$nm, and marks a change in
the scaling of the solvation free energy from being linear with solute
volume to being linear with exposed surface area. In bulk water, this
crossover provides a framework for understanding the assembly of small
species into a large aggregate.

Typical biological systems contain a high density of interfaces
including those of membranes and proteins, spanning the entire
spectrum from hydrophilic to hydrophobic. While water near hydrophilic
surfaces is bulk-like in many respects, water near hydrophobic
surfaces is different, akin to that near a liquid-vapor interface
\cite{stillinger,LCW,DC_nature05,mittal_pnas08,garde09pnas,ajp_jpcb2010}.
Here, we consider how these interfaces alter hydrophobic effects.
Specifically, to shed light on the thermodynamics of hydration at,
binding to, and assembly at interfaces, we study solutes with a range
of sizes at various self-assembled monolayer interfaces over a range
of temperatures using molecular simulations and theory.

Our principal results are that although the hydration thermodynamics
of hydrophobic solutes at hydrophilic surfaces is similar to that in
bulk, changing from entropic to enthalpic with increasing solute size,
it is enthalpic for solutes of all lengthscales near hydrophobic
surfaces.  Further, the driving force for
hydrophobically driven assembly in the vicinity of hydrophobic
surfaces is weaker than that in bulk, and decreases with increasing
temperature, in contrast to that in bulk. These results suggest that
hydrophobic surfaces will bind to and catalyze unfolding of proteins,
which we predict are relevant in the formation of amyloids and the
function of chaperonins.

\begin{figure}[bt]
\begin{center}\includegraphics[width=\textwidth]{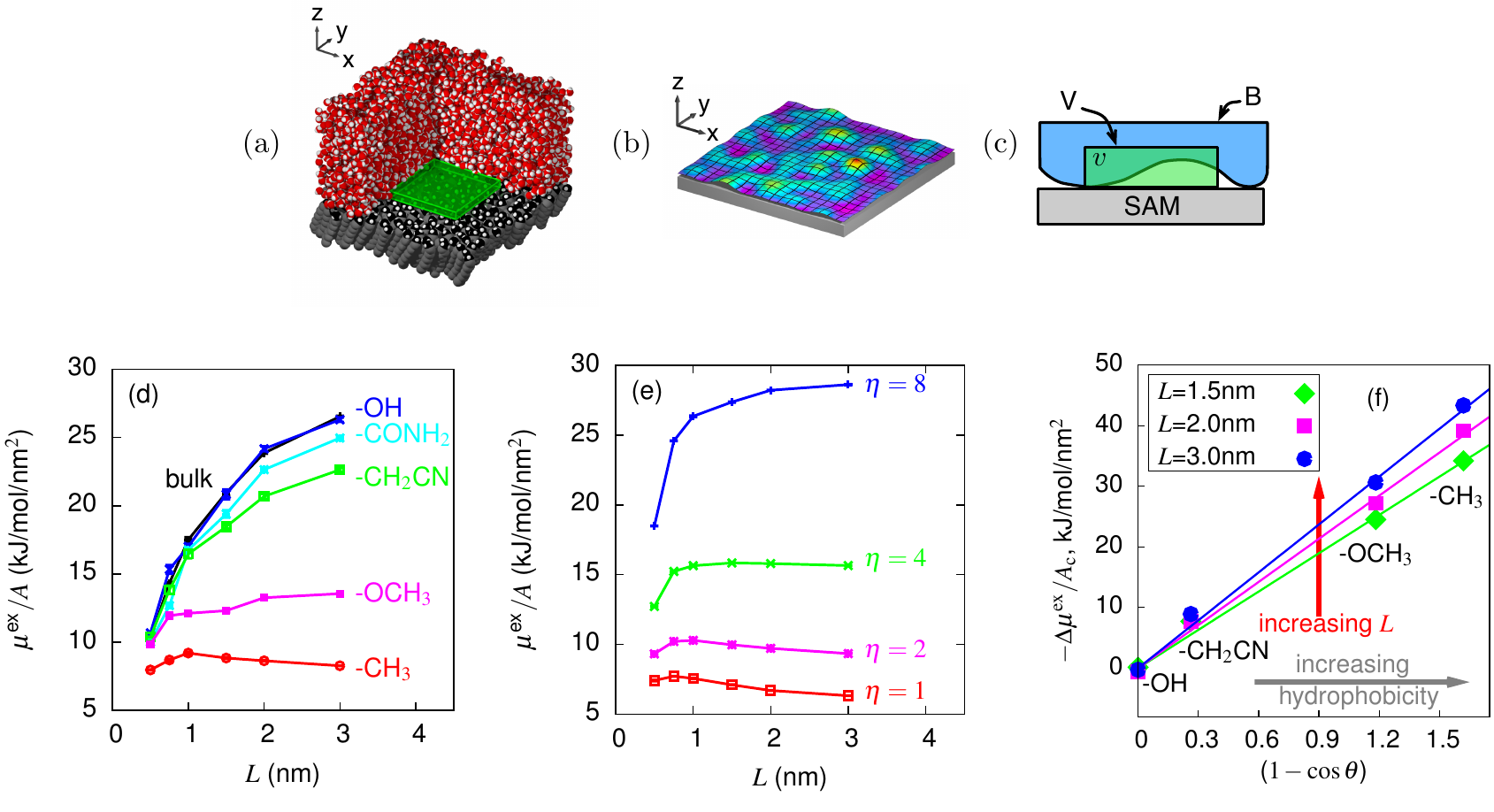}\end{center}
\caption{\label{fig:muex} Size dependent hydrophobic hydration at
interfaces. (a) A schematic of a cuboidal cavity (green) at the
SAM-water interface. The SAM head-groups (black and white), alkane
tails (grey), and water (red and white, partially cut out for clarity) are
shown. (b) A typical configuration of the model membrane, color-coded
by its distance from the model surface (grey). (c) Important volumes
in estimating the free energy~$\muex$ of emptying the probe volume~$V$
(green) using the theoretical model.  The region above the membrane is
the volume~$B$ (blue), and the intersection of $V$~and~$B$ is~$v$
(dark green).  (d) Lengthscale dependence of the cavity hydration free
energy per unit area, $\muex/A$, in bulk water and at interfaces, at
$T=300\,$K, obtained from MD simulations. (e) $\muex/A$, estimated
using the theoretical model, near surfaces with different attraction
strengths~$\eta$.  (f) Connecting the microscopic binding free energy
of a cavity to an interface, to the macroscopic surface
wettability. The $\cos\theta$ values were obtained from MD simulations
of a water droplet on SAM surfaces (\cite{garde09prl_HT}). Lines are
predictions using Eq.~\eqref{eqn:muexLigandYoung} with size dependent
$\gamma_{\rm LV}$ taken from (d).
}
\end{figure}

\subsection{Models}\ \\
{\bf Molecular simulations:} We simulate the solid-water interfaces of
self-assembled monolayers (SAMs) of surfactants [Fig.~\ref{fig:muex}(a)] with a range of
head-group chemistries, from hydrophobic (-CH$_3$) to hydrophilic
(-OH)~\cite{garde09pnas,garde09prl_HT}.
To study the size dependence of hydration at interfaces, we selected
cuboid shaped ($L \times L \times W$) cavities, with thickness
$W=0.3\,$nm, and side~$L$, varying from small values comparable to the
size of a water molecule to as large as ten times that size. Thicker
volumes will show qualitatively similar behavior, but will gradually
sample the ``bulk'' region away from the interface. \\ {\bf
Theoretical model:} To rationalize the simulation results and obtain
additional physical insights, we developed a model based on
Lum-Chandler-Weeks (LCW) theory~\cite{LCW}. LCW theory incorporates
the interplay between the small lengthscale gaussian density
fluctuations and the physics of interface formation relevant at larger
lengthscales, and captures the lengthscale dependence of hydrophobic
hydration in bulk water. Near hydrophobic surfaces, it predicts the
existence of a soft liquid-vapor-like interface, which has been
confirmed by
simulations~\cite{mittal_pnas08,garde09pnas,ajp_jpcb2010}.

We model this liquid-vapor-like interface near a hydrophobic surface,
as an elastic membrane [Fig.~\ref{fig:muex}(b)], whose energetics are
governed by its interfacial tension and the attractive interactions
with the surface.  The free energy of cavity hydration, $\mu^{\rm
ex}$, is related to the probability of spontaneously emptying out a
cavity shaped volume, $V$.  Such emptying can be conceptualized as a
two-step process in which interfacial fluctuations of the membrane can
empty out a large fraction of $V$ in the first step, with the
remaining volume $v$ emptied out via a density fluctuation
[Fig.~\ref{fig:muex}(c)]. When $v$ is small, the probability that it
contains $N$ waters is well-approximated by a Gaussian
\cite{information_theory,CrooksChandler1997,garde09pnas}.  The cost of
emptying $v$ can then be obtained from the average and the variance of
number of waters in $v$, which are evaluated by assuming that the
water density responds linearly to the surface-water adhesive
interactions.

We tune the strength of the model surface-water attraction,
$U(\mathbf{r})$, using a parameter $\eta$, where $\eta\approx 1$
corresponds to the hydrophobic $\text{-CH}_3$ SAM-like surface, with
higher values representing increasingly hydrophilic surfaces. The
representation of hydrophilic surfaces in our theoretical model lacks
the specific details of hydrogen bonding interactions (e.g.,
between the hydrophilic \hbox{-OH} SAM surface and water), so
comparisons between high-$\eta$ model surfaces and hydrophilic SAM
surfaces in simulations are qualitative in nature. Equations that put
the above model on a quantitative footing are given in the Appendix
and the details of its exact implementation are included as
Supplementary Information.

\subsection{Size-dependent hydrophobic hydration at, and binding 
to interfaces} Fig.~\ref{fig:muex}(d) shows the excess free energy,
$\muex$, to solvate a cuboidal cavity at temperature~$T=300\,$K,
divided by it's surface area ($A=2L^2+4LW$). $\muex/A$ can be thought
of as an effective surface tension of the cavity-water interface. In
bulk water, this value shows a gradual crossover with increasing $L$,
as expected \cite{LCW,garde05}.  Fig.~\ref{fig:muex}(d) also shows the
lengthscale dependence of $\muex/A$ for solvating cavities in
interfacial environments.  Near the hydrophilic OH-terminated SAM, the
behavior is similar to that in bulk water. However, with increasing
hydrophobicity of the interface, the size dependence of $\muex/A$
becomes less pronounced and is essentially absent near the -CH$_3$
surface, suggesting that hydration at hydrophobic surfaces is governed
by interfacial physics at all lengthscales.

Fig.~\ref{fig:muex}(e) shows the analogous solvation free energies
predicted using the theoretical model. 
The essential features of solvation next to the SAM surfaces are
captured well by this model.  This is particularly true for the
hydrophobic surfaces (with $\eta$~around~$1$), where the
potential~$U(\mathbf{r})$ closely mimics the effect of the real SAM on
the adjacent water, and the agreement between theory and simulation is
nearly quantitative.  For the more hydrophilic SAMs,
the comparison is qualitative, because the simple form for
$U(\mathbf{r})$ does not represent dipolar interactions well.

\begin{figure}
\begin{center}\includegraphics[width=\textwidth]{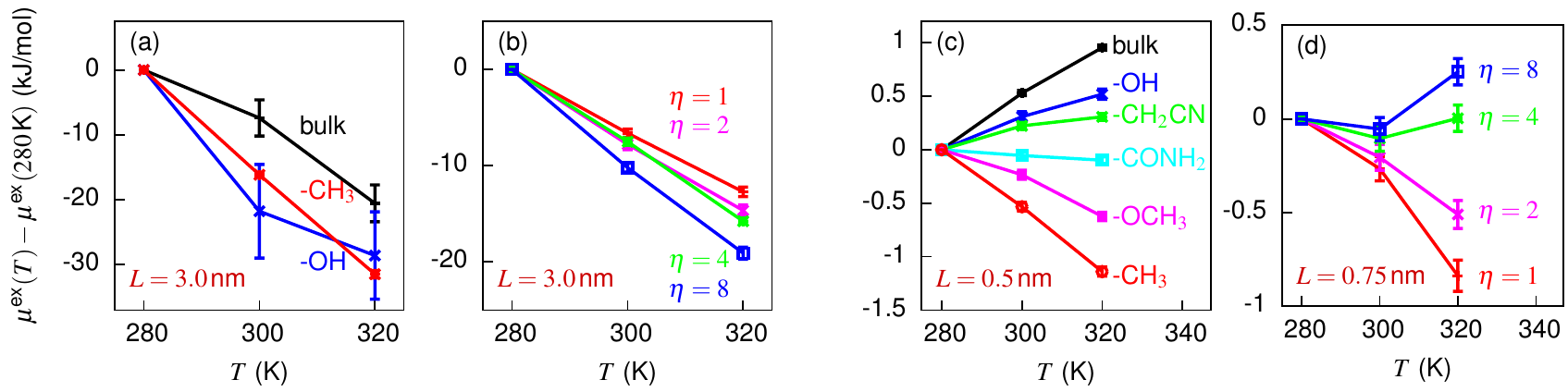}\end{center}
\caption{\label{fig:muexT}Temperature dependence of $\muex$ in bulk
water and at SAM-water interfaces for (a-b) large ($L=3.0\,$nm) and
(c-d) small ($L=0.5\,$nm, $L=0.75\,$nm) cavities, obtained from
simulations (a,c) and from the model (b,d) of
Eq.~\eqref{eqn:muexModel}.
}
\end{figure}

Fig.~\ref{fig:muex}(d) also indicates that $\muex$ becomes favorable
(smaller) with increasing surface hydrophobicity. The difference in
$\muex$ at an interface and in the bulk, $\Delta \muex= \muex_{\rm
int}-\muex_{\rm bulk}$, quantifies the hydration contribution to the
experimentally measurable free energy of binding of solutes to
interfaces.  Because the solvation of large solutes is governed by the
physics of interface formation, both in bulk and at the SAM surfaces,
we can approximate $\Delta \muex=A_{\rm c}(\gamma_{\rm SV}-\gamma_{\rm
SL}-\gamma_{\rm LV})$, where $A_{\rm c} = L^2$ is the cross-sectional
area, $\gamma$ is the surface tension, and subscripts SV, SL, and LV,
indicate solid-vapor, solid-liquid, and liquid-vapor interfaces,
respectively.  Using Young's equation, $\gamma_{\rm SV}=\gamma_{\rm
SL}+\gamma_{\rm LV}\cos\theta$, we rewrite
\begin{equation} 
\Delta \muex= -A_{\rm c}\gamma_{\rm LV}(1-\cos\theta),
\label{eqn:muexLigandYoung}
\end{equation}
where $\theta$ is the water droplet contact angle on the solid surface. 

Although Eq.~\eqref{eqn:muexLigandYoung} is strictly valid only for
macroscopic cavities, it can be applied to sufficiently large
microscopic cavitities with a lengthscale-dependent surface tension,
$\tilde\gamma_{\rm LV}$ ($\approx\muex_{\rm bulk}/A$). Indeed, lines
in Fig.~\ref{fig:muex}(f) predicted using
Eq.~\eqref{eqn:muexLigandYoung} are in excellent agreement with
simulation data, and indicate that the strength of binding increases
with surface hydrophobicity as well as with solute size. These results
establish a connection between the microscopic solute binding free
energies to interfaces and the macroscopic wetting properties of those
interfaces.  This connection provides an approach to characterize the
hydrophobicity of topographically and chemically complex interfaces,
such as those of proteins
\cite{acharya:faraday,Giovambattista:PNAS:08}.

\subsection{Temperature dependence of hydration at interfaces}
The differences between the cavity hydration at interfaces and in bulk
are highlighted most clearly in the $T$-dependence of~$\muex$, which
characterizes the entropic and enthalpic contributions to the free
energy. For small solutes in bulk, the entropy of hydration is known
to be large and negative \cite{garde96,HC_temp}, which reflects the
reduced configurational space available to the surrounding water
molecules.  In contrast, for large solutes the entropy of hydration is
expected to be positive, consistent with the temperature dependence of
the liquid-vapor surface tension
\cite{water_gamma_vs_T}. Fig.~\ref{fig:muexT}(a) shows that $\muex$ of
large cuboidal cavities ($L=3\,$nm) in bulk water indeed decreases
with increasing temperature, although the corresponding hydration
entropy per unit surface area ($25\,$J/mol/K/nm$^2$) is lower than
that expected from the temperature derivative of surface tension of
water (about~$90\,$J/mol/K/nm$^2$~\cite{water_gamma_vs_T}).  We note
that solvation entropies in SPC/E water obtained using NPT ensemble MD
simulations are known to be smaller than experimental values by
about~$20$\%~\cite{garde08salt}. Additionally, the cavity-water
surface tension and its temperature derivative for these nanoscopic
cavities are expected to be smaller than the corresponding macroscopic
values \cite{ajp_jpcb2010}.

Fig.~\ref{fig:muexT}(a) also shows that for large cuboidal cavities
($L=3\,$nm), $\muex$ decreases with increasing temperature not only in
bulk water and near the hydrophilic (-OH) surface, but also near the
hydrophobic (-CH$_3$) surface, indicating a positive entropy of cavity
formation. Thus, in all three systems, the thermodynamics of hydration
of large cavities is governed by interfacial physics.  Although the
values of $\muex(L=3\,\text{nm})$ at $300\,$K are rather large
($569\,$kJ/mol in bulk water, $565\,$kJ/mol at the -OH interface, and
$167\,$kJ/mol at the -CH$_3$ interface), their variation with
temperature shown in Fig.~\ref{fig:muexT}(a) is similar in bulk and at
interfaces.

Fig.~\ref{fig:muexT}(b) shows that this same phenomenology is captured
nearly quantitatively by the theoretical model.  In the model, the
cavity hydration free energies have large but athermal contributions
from the attractions between water and the model surface.  The main
temperature-dependent contribution to $\muex$ is the cost to deform
the liquid-vapor-like interface near the surface to accommodate the large
cavity.  Since the necessary deformation is similar, regardless of the
hydrophobicity of the surface, the variation of $\muex$ with
temperature is similar as well.

Fig.~\ref{fig:muexT}(c) shows the temperature dependence of $\muex$
for small cavities ($L=0.5\,$nm) in bulk and at SAM-water surfaces.
In bulk water, $\muex$ increases with temperature, and yields an
entropy of hydration of roughly $-25\,$J/mol/K, characteristic of
small lengthscale hydrophobic hydration. This negative value is
consistent with those calculated for spherical solutes of a similar
volume \cite{garde96}. With increasing hydrophobicity, the slope of
the $\muex$ vs $T$ curve decreases and becomes negative, indicating a
positive entropy of cavity formation near sufficiently hydrophobic
surfaces. Near the most hydrophobic surface (-CH$_3$), the entropy of
hydration of this small cavity is $+30\,$J/mol/K.

Fig.~\ref{fig:muexT}(d) shows that the same phenomenon is recovered by
the theoretical model, though the correspondence is clearest at a
slightly larger cavity size ($L=0.75\,$nm).  Near hydrophilic model
surfaces, the interface is pulled close to the surface by a strong
attraction, so it is costly to deform it.  As a result, the cavity is
emptied through bulk-like spontaneous density fluctuations that result
in a negative entropy of hydration of small cavities. In contrast,
near a hydrophobic surface, the interface is easy to deform, which
provides an additional mechanism for creating cavities.  In fact, this
mechanism dominates near sufficiently hydrophobic surfaces, and since
the surface tension of water decreases with increasing temperature, so
does $\muex$.  Hence, even small cavities have a positive entropy of
hydration near hydrophobic surfaces.  The continuous spectrum of
negative to positive solvation entropies observed in
Figs.~\ref{fig:muexT}(c-d) is thus revealed to be a direct consequence
of the balance between bulk-like water density fluctuations and
liquid-vapor-like interfacial fluctuations.

\begin{figure}[bth]
\begin{center}
\includegraphics[width=4in]{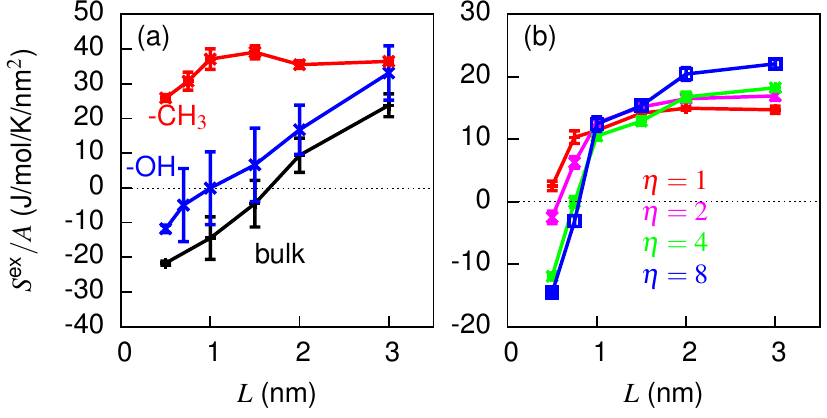}
\end{center}
\vspace*{-0.3in}
\caption{\label{fig:SexA} Lengthscale dependence of the excess
solvation entropy per unit surface area for (a) cavities in bulk water
and at the -CH$_3$ and -OH SAM-water interfaces, and (b) cavities in
the model of Eq.~\eqref{eqn:muexModel} near surfaces of different
attraction strengths,~$\eta$.}
\end{figure}

Fig.~\ref{fig:SexA}(a) shows that near the hydrophobic
CH$_3$-terminated SAM, cavity hydration entropies per unit area,
$S^{\rm ex}/A$, are positive and essentially constant (about
$30\,$J/mol/K/nm$^2$) over a broad range of cavity sizes.  In
contrast, in bulk water, $S^{\rm ex}/A$ depends on $L$, and changes
from large negative to positive values with increasing $L$.  The
lengthscale at which entropy crosses zero, $L_{\text{S}}$, can serve
as a thermodynamic crossover length.  In bulk water,
$L_{\text{S}}\approx 1.8\pm0.2\,$nm. The behavior of $S^{\rm ex}/A$ is
qualitatively similar at the -OH surface, with $L_{\rm S}\approx
1.3\pm0.4\,$nm. Although the numerical value of $L_{\text{S}}$ may
depend on the shape of the cavity and on solute-water attractions for
non-idealized hydrophobes, the trend in entropy should not.

Fig.~\ref{fig:SexA}(b) shows that our implementation of LCW
ideas recovers many of the observed trends, with solvation entropy
being everywhere positive for the smallest attraction strength~$\eta$,
and a thermodynamic crossover length of just under~$1\,$nm emerging
for the more hydrophilic model surfaces, similar to that in bulk
water.  Nevertheless, the agreement between
Figs. \ref{fig:SexA}(a)~and~(b) is somewhat qualitative, mostly as a
result of the crude form of $U(\mathbf{r})$ used to model hydrophilic
surfaces.

\subsection{Thermodynamics of binding to, and assembly at 
hydrophobic surfaces} 
\begin{figure}[tbh]
\begin{center}
\includegraphics[angle=0,width=3.5in]{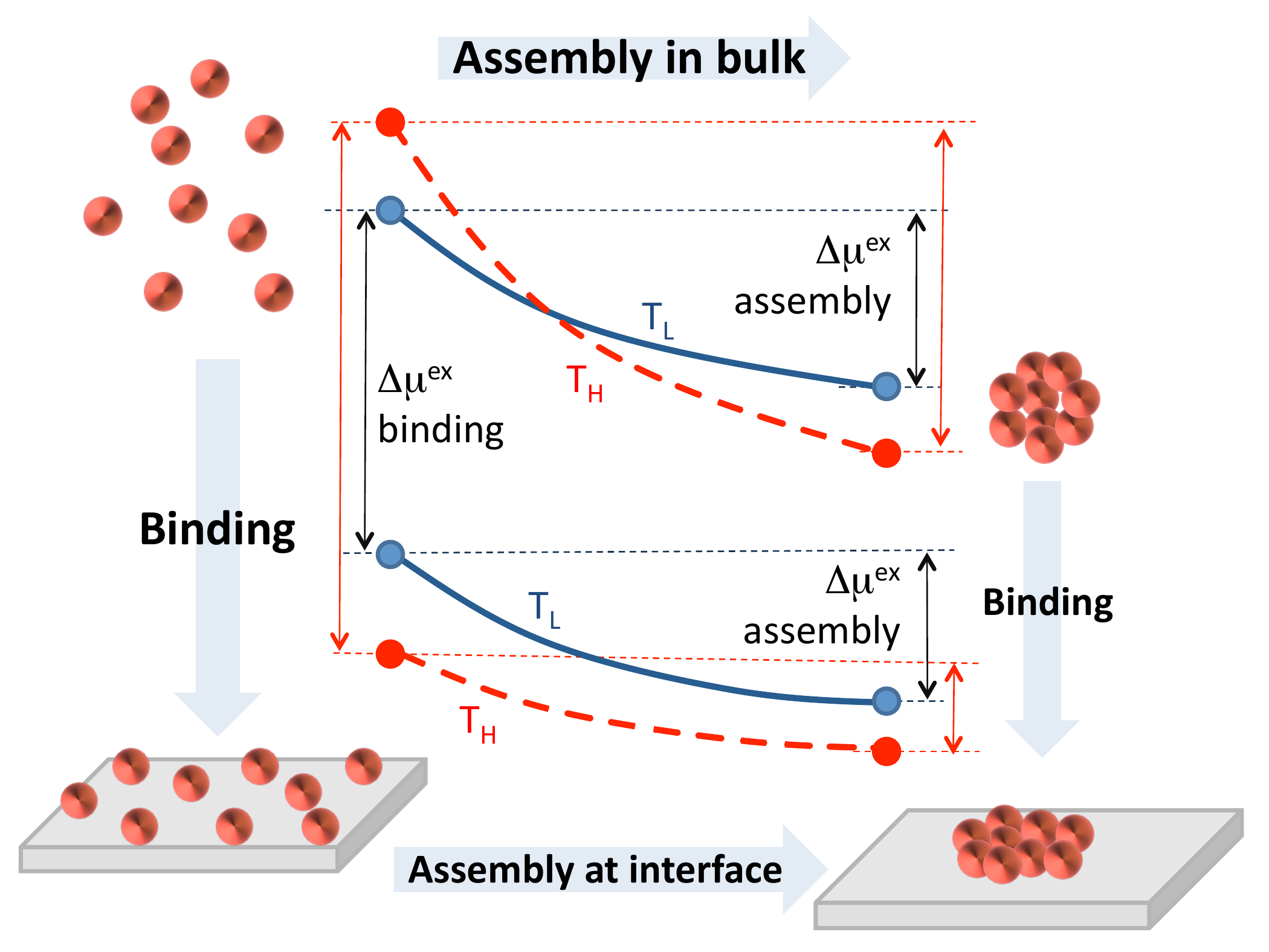}
\end{center}
\caption{\label{fig:schematic} Schematic illustrating the
thermodynamics of binding and assembly. The points represent free
energies of solvating small objects individually (left) and in the
assembled state (right), in bulk (top) and at a hydrophobic interface
(bottom), at a lower (blue, $T_{\rm L}$) and a higher (red, $T_{\rm
H}$) temperature near ambient conditions. Assembly: The driving force
for assembly at hydrophobic interfaces is smaller than that in
bulk. It is enthalpic and decreases with increasing temperature,
unlike in bulk. Binding: The driving force for binding small objects
to a hydrophobic surface increases with temperature, so it is
entropic.  For large objects, it is enthalpic.
}
\end{figure}

In the preceding sections, we have examined the
hydration behavior of single, isolated, idealized cavities near flat
surfaces and in the bulk.  We now consider the consequences of our
observations on hydrophobically driven binding and assembly,
summarized schematically in Fig.~\ref{fig:schematic}.

Fig.~\ref{fig:schematic} indicates that while the binding of both
small and large solutes (or aggregates) to hydrophobic surfaces is
highly favorable, their thermodynamic signatures are different.
Binding of small solutes is entropic and becomes more favorable with
increasing temperature, whereas binding of large solutes is enthalpic
and depends only weakly on temperature.

Fig.~\ref{fig:schematic} also highlights the differences in the
thermodynamics of hydrophobically driven assembly at interfaces and in
bulk, inferred from our lengthscale dependence studies. In bulk, the
solvation of many small, isolated hydrophobes scales as their excluded
volume.  Accommodating small species inside the existing
hydrogen-bonding network of water imposes an entropic cost, so the
solvation free energy increases with increasing temperature.  When
several small hydrophobes come together, water instead hydrates the
aggregate by surrounding it with a liquid-vapor-like interface. The
corresponding solvation free energy scales as the surface area and
decreases with increasing temperature.

Thus, the driving force for assembly of $n$ small solutes (each of
surface area $A_1$, volume $v_1$ and solvation free energy of
$\mu_{1,\mathrm{bulk}}^{\mathrm{ex}}$) into a large aggregate (with
surface area $A_n$ and volume $nv_1$) in bulk water is
well-approximated by
\begin{equation}
\Delta\muex_{\mathrm{bulk}} = \gamma_{\mathrm{bulk}} A_n -
n\mu_{1,\mathrm{bulk}}^{\mathrm{ex}} =
[Cn^{-1/3}-1]n\mu_{1,\mathrm{bulk}}^{\mathrm{ex}},
\label{eqn:muexBulk}
\end{equation}
where $C\sim(\gamma_{\mathrm{bulk}}
v_1^{2/3}/\mu_{1,\mathrm{bulk}}^{\mathrm{ex}})$ and
$\gamma_{\text{bulk}}$ is a curvature-corrected effective surface
tension [top curve of Fig.~\ref{fig:muex}(d)].  As the surface tension
decreases with increasing temperature, so does the free energy to
hydrate nanometer-sized aggregates. However, the free energy to
individually hydrate the small solutes increases with temperature,
resulting in a larger driving force for assembly.  Conversely, while
the driving force for assembly, $\Delta\muex_{\text{bulk}}$, is large
and negative (favorable) at ambient conditions, it decreases in
magnitude with decreasing temperature [upper portion of
Fig.~\ref{fig:schematic}], and can even change sign at a sufficiently
low temperature.  When adapted to particular systems,
Eq.~\eqref{eqn:muexBulk} can, with remarkable accuracy, explain
complex solvation phenomena like the temperature-dependent aggregation
behavior of micelles~\cite{MaibaumDinnerChandler2004} and the cold
denaturation of proteins~\cite{kauzmann}.

In the presence of a hydrophobic surface, on the other hand, we have
found that interfacial physics dominates at all lengthscales
[Fig.~\ref{fig:muexT}(a-d) and Fig.~\ref{fig:SexA}(a-b)].  As a
result, the driving force for assembly at interfaces,
$\Delta\muex_{\mathrm{int}}$, does not scale as in
Eq.~\eqref{eqn:muexBulk}, but is instead given by
\begin{equation}
\Delta\muex_{\mathrm{int}} = \gamma_{\mathrm{int}}(A_n - nA_1) \sim
[n^{-1/2}-1]n\mu_{1,\mathrm{int}}^{\mathrm{ex}},
\label{eqn:muexInt}
\end{equation}
where $\gamma_{\mathrm{int}}$ is the effective surface tension at the
interface, [the lower curves of Figs.~\ref{fig:muex}(d-e)]. Since
$\gamma_{\mathrm{int}}$ decreases with increasing temperature
[Figs.~\ref{fig:muexT}(a-d)], so does the hydration contribution to the
driving force for assembly at a hydrophobic surface, in contrast to that in bulk.

The free energy barrier between disperse and assembled states is also
expected to be very different in bulk and near hydrophobic surfaces.
In bulk, the dispersed state has no liquid-vapor-like interface
whereas the assembled state does.  The transition state consists of a
critical nucleus of hydrophobic particles that nucleates the
liquid-vapor-like interface.  The nucleation barrier can be high, and
dominates the kinetics of hydrophobic collapse of idealized
hydrophobic polymers~\cite{tenWoldeChandler2002,tommy,pgd:jpcb:2009}
and plates~\cite{HuangZhouBerne2005}.  In contrast, we expect
aggregation near hydrophobic surfaces to be nearly barrierless, since
an existing liquid-vapor-like interface is deformed continuously
between the disperse and assembled states.

Finally, and most importantly, we find that for large aggregates, the
driving force of assembly is weaker near interfaces than in bulk.  In
the limit of large $n$, the terms $n\muex_1$ dominate both at
interfaces and in bulk (Eqns.~\eqref{eqn:muexBulk}
and~\eqref{eqn:muexInt}), and the results in Fig.~\ref{fig:muex}(d)
show that $\muex_{1,\text{int}} < \muex_{1,\text{bulk}}$.

The nontrivial behavior of the driving forces and barriers to
assembly at interfaces should be relevant in biological systems where
hydrophobicity plays an important role. Experiments have shown that
hydrophobic surfaces bind and facilitate the unfolding of proteins,
including those that form amyloids~\cite{Radke,Belfort,pomes}. Our
results shed light on these phenomena and suggest that large
hydrophobic surfaces may generically serve as catalysts for unfolding
proteins \cite{Berne:10:BJ}, via solvent-mediated interactions.
Indeed, simulations show that the binding of model hydrophobic
polymers to hydrophobic surfaces is accompanied by a conformational
rearrangement from globular to pancake-like
structures~\cite{SJ_jpcb09}. Such conformations can further assemble
into secondary structures, such as
$\beta$-sheets~\cite{shea08,Berne:10:BJ,Belfort,pomes}, and we predict
that the solvent contribution to this assembly at the hydrophobic
surface will be governed by interfacial physics. This implies that
manipulating the liquid-vapor surface tension, either by changing the
temperature or by adding salts or co-solutes, will allow one to
manipulate the driving force for assembly.

We further speculate that the catalysis of unfolding by hydrophobic
surfaces may play a role in chaperonin function~\cite{Chaperonins}.
The interior walls of chaperonins in the open conformation are
hydrophobic and can bind misfolded proteins, whereupon their unfolding
is catalyzed \cite{Pande,JEShea_rev}. Subsequent ATP-driven
conformational changes render the chaperonin walls hydrophilic
\cite{Pande,Chaperonins}.  As a result, the unfolded protein is
released from the wall, as the free energy for a hydrophobe to bind to
a hydrophilic surface is much lower than that to bind to a hydrophobic
one [Fig.~\ref{fig:muex}(d)].

Our results also provide insights into the interactions between
biomolecules and nonbiological hydrophobic surfaces, such as those of
graphite and of certain metals, which have been shown to bind and
unfold proteins \cite{Marchin:Langmuir:2003,Belfort:Langmuir:2011}.
Such interactions are of interest in diverse applications including
nano-toxicology \cite{Tian:2006} and biofouling
\cite{Belfort:Langmuir:2011}.

Collectively, our findings highlight that assembly near hydrophobic
surfaces is different from assembly in bulk and near hydrophilic
surfaces. Experimental measurements of the thermodynamics of protein
folding have been performed primarily in bulk water
\cite{privalov}. Although many experiments have probed how interfaces
affect protein folding, structure and function~\cite{Radke,dordick},
to the best of our knowledge, there are no temperature-dependent
thermodynamic measurements of self-assembly at interfaces. We hope
that our results will motivate such measurements.

\appendix
\section*{Appendix}

\begin{figure}[tbh]
\begin{center}
\includegraphics[width=4in]{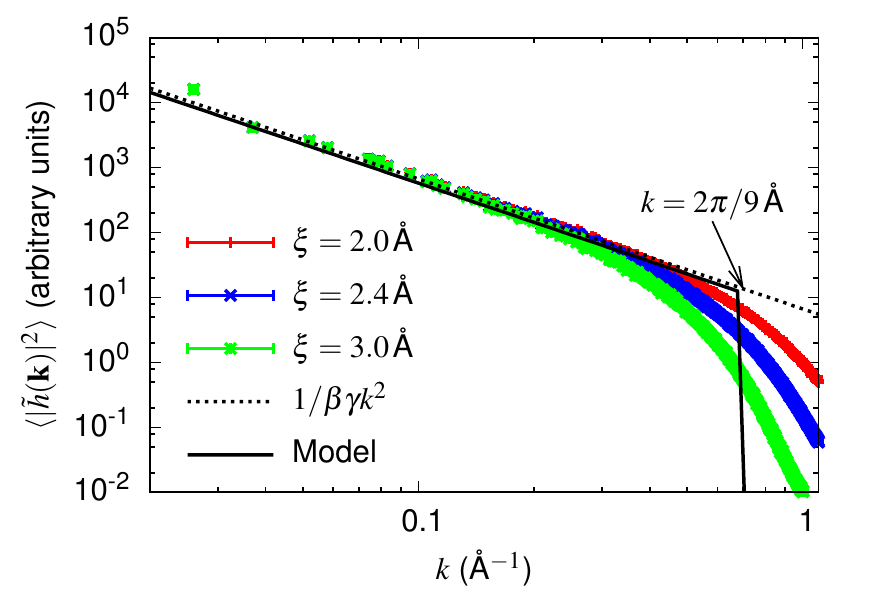}\end{center}
\vspace*{-0.3in}
\caption{
\label{fig:powerSpectrum}
Power spectrum of the instantaneous liquid-vapor interface
at~$T=300\,$K.  A liquid-vapor interface was simulated using a
$24\times24\times3\,$nm$^3$ slab of SPC/E water in a periodic box of
size $24\times24\times9\,$nm$^3$ and the instantaneous interface
configuration,~$h(x,y)$, and its Fourier transform,
$\hTwiddle(\mathbf{k})$, were evaluated as in
Ref. \cite{WillardChandler2010}.  The power spectrum of our simulated
instantaneous interface is good agreement with the capillary-wave
theory prediction ($\langle |\hTwiddle(
\mathbf{k})|^2\rangle\sim1/\beta\gamma k^2$) for wavevectors smaller
than $\sim 2\pi / 9\,$\AA. For larger wavevectors, the power spectrum
is sensitive to molecular detail, {\it i.e.}, the coarse-graining
length~$\xi$ used to define the intrinsic interface, as expected
~\cite{SedlmeierEtAl2009}.  Fitting the $\xi = 2.0\,$\AA\ data in the
range $0.01\,\text{\AA}^{-1} < k < 0.3\,\text{\AA}^{-1}$ yields
$\gamma = 62.0\pm0.5\,$mJ/m$^2$, in reasonable agreement with the
experimental value of $72\,$mJ/m$^2$ and some simulated values of the
SPC/E surface tension ({\it e.g.},
$63.6\pm1.5\,$mJ/m$^2$~\cite{VegaMiguel2007}), but not others ({\it
e.g.}, $52.9\,$mJ/m$^2$ ~\cite{SedlmeierEtAl2009}).
}
\end{figure}

{\bf Simulation details:} Our simulation setup and force fields are
similar to that described in
Refs. \cite{garde09prl_HT,garde09pnas}. Simulations were performed in
the NVT ensemble with a periodic box
($7\,$nm$\times7\,$nm$\times9\,$nm) that has a buffering liquid-vapor
interface at the top of the box, for reasons explained in
Ref.~\cite{ajp_jpcb2010}. It has been shown that free energies
obtained in the above ensemble are indistinguishable from those
obtained in the NPT ensemble at a pressure of
$1\,$bar~\cite{ajp_jstatphys}.  We have chosen the SPC/E model of
water~\cite{spce} since it adequately captures experimentally
known features of water, such as surface tension, compressibility, and
local tetrahedral order, that play important roles in the
hydrophobic effect~\cite{DC_nature05}.  Electrostatic interactions
were calculated using the particle mesh Ewald method \cite{PME}, and
bonds in water were constrained using SHAKE \cite{SHAKE}. Solvation
free energies were calculated using test particle insertions
\cite{Widom_jcp63} for smaller cavities ($L<1\,$nm), and the indirect
umbrella sampling (INDUS) method \cite{ajp_jpcb2010,ajp_jstatphys} for
larger cavities.
\\
{\bf Theoretical Model:} We model the liquid-vapor-like interface near
hydrophobic surfaces as a periodic elastic membrane, $z = h(x,y)$,
with an associated Hamiltonian, $H[h(x,y)]$:
\begin{equation}
H[h(x,y)] = \int_{x,y} \left[ \frac{\gamma}{2} |\mathbf{\nabla}
h(x,y)|^2 + \int_{z=h(x,y)}^\infty \rho_\ell U(\mathbf{r})\right].
\label{eqn:Ham}
\end{equation}
Here, $\gamma$ is the experimental liquid-vapor surface tension of
water, $\rho_\ell$ is the bulk water density, and $U(\mathbf{r})$ is
the interaction potential between the model surface and a water
molecule at position $\mathbf{r} = (x,y,z)$.  The square-gradient term
in Eq.~\eqref{eqn:Ham} accurately captures the energetics of
interfacial capillary waves only for wavelengths larger than atomic
dimensions (Fig. \ref{fig:powerSpectrum}), so we restrict $h(x,y)$ to
contain modes with wavevectors below $2\pi/9\,$\AA.  At any instant in
time, part of $V$ can be empty due to an interfacial fluctuation.  The
number of waters in the remaining volume, $v$, fluctuates, and we
denote by $P_v(N)$ the probability that $v$ contains $N$ waters.  We
thus estimate the free energy for emptying~$V$ completely to be
\begin{equation}
\muex(V) = -\kB T \ln \int \mathcal{D}h\,Z^{-1} e^{-\beta H[h(x,y)]}
P_v(0),
\label{eqn:muexModel}
\end{equation}
where $Z=\int \mathcal{D}h\,\exp\{-\beta H[h(x,y)]\}$ is the partition
function of the membrane. The volume~$v$ depends on the interfacial
configuration~$h(x,y)$, {\it i.e.}, $v = v[h(x,y)]$.

It is known that~$P_v(N)$ is well-approximated by a Gaussian when
$v$~is small~\cite{information_theory,
CrooksChandler1997,garde09pnas}.  If water were far from liquid-vapor
coexistence, then $P_v(N)$ would also be close to Gaussian for
arbitrarily large~$v$.  The fact that water at ambient conditions is
near liquid-vapor coexistence, and that there is a liquid-vapor-like interface
near the SAM, is captured by the additional interfacial energy
factor~$Z^{-1} \exp\{-\beta H[h(x,y)]\}$ in Eq.~\eqref{eqn:muexModel}.
The net result is that the thermal average of
Eq.~\eqref{eqn:muexModel} is dominated by interface configurations
where $v$ is small, so that even at ambient conditions, we can
approximate
$$ P_v(N) \approx (2\pi \sigma_v)^{-1/2} \exp\left[ -(N - \langle N
\rangle_v)^2 / 2\sigma_v \right],
$$ where $\langle N \rangle_v$ is the average number of waters in~$v$
and $\sigma_v = \langle (\delta N)^2 \rangle_v$ is the variance.  We
estimate these by noting that the solvent density responds linearly to
the attractive potential,~$U(\mathbf{r})$, in the volume occupied by
the water,~$B$, depicted in
Fig.~\ref{fig:muex}(c)~\cite{gaussian_ft,information_theory,LLCW}.
Hence,
\begin{align*}
\langle N\rangle_v &\approx \rho_\ell v - \int_{\mathbf{r}\in v}
\int_{\mathbf{r}'\in B}\, \chi(\mathbf{r},\mathbf{r}') \beta
U(\mathbf{r}'),\\ 
\sigma_v &\approx \int_{\mathbf{r}\in v}
\int_{\mathbf{r}'\in v} \chi(\mathbf{r},\mathbf{r}'), \text{where} \\
\chi(\mathbf{r},\mathbf{r}') &= \rho_\ell \delta(\mathbf{r} -
\mathbf{r}') + \rho_\ell^2 [g(|\mathbf{r} - \mathbf{r}'|) - 1].
\end{align*} 
Here, $g(r)$ is the oxygen-oxygen radial distribution function of
water~\cite{NartenLevy1971}.

The surface--water interaction is modeled by a
potential,~$U(\mathbf{r})$, that closely mimics the attractive
potential exterted by the $\text{-CH}_3$~SAM on water:
$$ U(\mathbf{r}) = U_{\text{wall}}(\mathbf{r}) + \eta
U_{\text{head}}(\mathbf{r}) + U_{\text{tail}}(\mathbf{r}).
$$ The first term, $U_{\text{wall}}(\mathbf{\mathbf{r}})$, is a
sharply repulsive potential in the region $z < R_0$ that captures the
hard-core exclusion of a plane of head groups at~$z=0$ with
hard-sphere radius~$R_0$.  The second term,
$U_{\text{head}}(\mathbf{r})$, captures the head group--water
interaction, modeled as a plane of OPLS/UA CH$_3$ Lennard-Jones (LJ)
interaction sites~\cite{JorgensenMaduraSwenson1984} of area
density~$\mu_{\text{head}}$ at $z = 0$, and is scaled by~$\eta$.  The
final term, $U_{\text{tail}}(\mathbf{r})$, similarly captures the
alkane tail--water interaction, modeled as a uniform half-space of
OPLS/UA CH$_2$ LJ interaction sites of volume
density~$\rho_{\text{tail}}$ at a distance~$\zeta$ below the head
groups.  The parameters $R_0$, $\zeta$,
$\mu_{\text{head}}$~and~$\rho_{\text{tail}}$ are dictated by the
geometry of the SAM (See SI for details).  

\begin{acknowledgments}
The authors would like to thank Steve Granick, Bruce Berne and Frank Stillinger
for providing helpful comments on an earlier draft. AJP~and~PV were supported by NIH Grant No. R01-GM078102-04.
SG~gratefully acknowledges financial support of the NSF-NSEC
(DMR-0642573) grant.  DC~was supported by the Director, Office of
Science, Office of Basic Energy Sciences, Materials Sciences and
Engineering Division and Chemical Sciences, Geosciences, and
Biosciences Division of the U.S. Department of Energy under Contract
No. DE-AC02-05CH11231.
\end{acknowledgments}

\section*{Supplementary Information}

Here, we describe the details of the solvation model used in the main text.

\subsection{Interface description}

We describe the liquid-vapor-like interface next to the model surface by a periodic height function~$h(\va)$, with $\va = (x,y)$ and $-D/2 \leq x,y < D/2$.  This function is sampled discretely at a resolution~$\Delta$, at points satisfying
$$
\va = (n_x \Delta, n_y \Delta),\qquad -\frac{D}{2\Delta} \leq n_x, n_y < \frac{D}{2\Delta}.
$$
This results in $N^2$ discrete sampling points~$\{\va\}$, with $N = D/\Delta$.  In the following, sums over $\va$ denote sums over these $N^2$ sampling points.  We have used $D = 60\,$\AA\ and $\Delta = 1\,$\AA.

The discrete variables~$\{h_\va\}$ represent the interface height at each sample point~$\va$, so that
$$
h_\va = h(\va),\quad\text{for $\va = (n_x\Delta, n_y\Delta)$.}
$$
This notation clearly distinguishes between the $N^2$ height variables $h_\va$ and the continuous height function $h(\va)$ that they represent.

The discrete Fourier transform of~$\{h_\va\}$ is denoted by $\{\hTwiddle_\vk\}$, and is defined at wavevectors~$\vk = (2\pi/L) ( m_x, m_y )$, with $-N/2 \leq m_x, m_y < N/2$.  We use the symmetric normalization convention throughout for Fourier transforms.

\subsection{Energetics}

The essential property of the liquid-vapor-like interface is its surface tension, which results in the following capillary-wave Hamiltonian~\cite{BuffLovettStillinger1965} for a free interface,
$$
H_0[\{h_\va\}] \approx \frac{\gamma\Delta^2}{2} \sum_\va | \nabla h_\va |^2 \approx \frac{\gamma\Delta^2}{2} \sum_\vk k^2 |\hTwiddle_\vk|^2,
$$
where $\nabla h_\va$ is a finite-difference approximation to~$\nabla h(\va)$.

Using an appropriate definition of an instantaneous water-vapor interface~\cite{WillardChandler2010}, the power spectrum of capillary waves in SPC/E water has been found to agree with the spectrum predicted by the above Hamiltonian for wavevectors smaller than about $2\pi/\ell$, but is substantially lower for higher wavevectors (Fig.~6 of main text).  This result is consistent with the liquid-vapor-like interfaces being sensitive to molecular detail at high wavevectors~\cite{SedlmeierEtAl2009}.  At $T=300\,K$, we have found that
$
\ell \approx 9\,\text{\AA}.
$
We thus constrain all Fourier components $\hTwiddle_\vk$ to be zero for high $\vk$, i.e.
\begin{equation}
\hTwiddle_\vk = 0,\qquad |\vk| > 2\pi/\ell.
\label{eqn:noHighKs}
\end{equation}

In our model, the liquid-vapor-like interface interacts with a model surface via a potential that depends on~$\{h_\va\}$.  As discussed below, it is also convenient to introduce additional umbrella potentials to aid in sampling.  The Hamiltonian of the interface subject to this additional potential energy~$H'[\{h_\va\}]$ is
\begin{equation}
H[\{h_\va\}] = \frac{\gamma\Delta^2}{2} \sum_\vk k^2 |\hTwiddle_\vk|^2 + H'[\{h_\va\}].
\label{eqn:discreteHam}
\end{equation}
When expressed as a function of the Fourier components~$\{\hTwiddle_\vk\}$, we denote the Hamiltonian by $\HTwiddle[\{\hTwiddle_\vk\}]$ and the external potential by $\HpTwiddle[\{\hTwiddle_\vk\}]$, so that
$$
\HTwiddle[\{\hTwiddle_\vk\}] = \frac{\gamma\Delta^2}{2} \sum_\vk k^2 |\hTwiddle_\vk|^2 + \HpTwiddle[\{\hTwiddle_\vk\}].
$$


\subsection{Dynamics}

We calculate thermal averages of interface configurations by introducing a fictitious Langevin dynamics and replacing thermal averages by trajectory averages.  We first assign a mass per unit area~$\mu$ to the interface.  The Lagrangian in real space is
$$
L[\{h_\va,\hDot_\va\}] = \frac{\mu\Delta^2}{2} \sum_\va \hDot_\va^2 - H[\{h_\va\}].
$$
The corresponding Lagrangian in Fourier space is
$$
{\tilde L}[\{\hTwiddle_\vk,\hDotTwiddle_\vk\}] = \frac{\mu\Delta^2}{2} \sum_\vk |\hDotTwiddle_\vk|^2 - \HTwiddle[\{\hTwiddle_\vk\}].
$$
Since all $h_\va$ are real, the amplitudes of modes $\vk$~and~$-\vk$ are related,
$
\hTwiddle_\vk = \hTwiddle^*_{-\vk}.
$
Taking this constraint and Equation~\eqref{eqn:noHighKs} into account, the Euler-Lagrange equations yield equations of motion in Fourier space.  To thermostat each mode, we add Langevin damping and noise terms.  The final equation of motion has the form
\begin{equation}
\mu\Delta^2\hDDotTwiddle_\vk = -\gamma\Delta^2 |\vk|^2 \hTwiddle_\vk - \frac{\del \HpTwiddle[\{\hTwiddle_\vk\}]}{\del \hTwiddle_\vk} - \eta \hDotTwiddle_\vk + \xiTwiddle_\vk(t),\qquad(|\vk| < 2\pi/\ell),
\label{eqn:EoMs}
\end{equation}
The Langevin damping constant~$\eta$ is chosen to decorrelate momenta over a timescale~$\tau$, so 
$
\eta = \mu\Delta^2 / \tau.
$
The zero-mean Gaussian noise terms~$\{\xiTwiddle_\vk(t)\}$ have variance such that
$$
\langle \xiTwiddle^*_\vk(t) \xiTwiddle_\vk(t') \rangle = 2\eta \kB T \delta( t - t').
$$
As with $\hTwiddle_\vk$, $\xiTwiddle_\vk$  satisfy the related constraint $\xiTwiddle_\vk = \xiTwiddle^*_{-\vk}$.  Hence, for $\vk = \vec 0$, the noise is purely real and its variance is twice that of the real and imaginary components of all other modes
\footnote{The constraint on the magnitude of $\vk$ ensures that no Nyquist modes, i.e., modes with $k_x$~or~$k_y$ equal to $\pm\pi/D$, are ever excited.  If they were included, these modes would also be purely real, and the variance of the real component of their noise terms would likewise be twice that of the real component of the interior modes.}.

We propagate these equations of motion using the Velocity Verlet algorithm.  At each force evaluation, we use a Fast Fourier Transform (FFT) to calculate $\{h_\va\}$ from $\{\hTwiddle_\vk\}$.  We then calculate $\del H'[\{h_\va\}] / \del h_\va$ in real space and perform an inverse FFT to obtain the force $\del \HpTwiddle[\{\hTwiddle_\vk\}] / \del \hTwiddle_\vk$ on mode~$\hTwiddle_\vk$ due to $H'[\{h_\va\}]$.  We then add the forces due to surface tension, Langevin damping and thermal noise, as in Eq.~\eqref{eqn:EoMs}.

For the Velocity Verlet algorithm to be stable, we choose a timestep equal to~$1/20^{\text{th}}$ of the typical timescale of the highest-frequency mode of the free interface,
$
\Delta t = \frac{1}{20} \sqrt{\mu\ell^2/\gamma}.
$
To equilibrate the system quickly but still permit natural oscillations, we choose the Langevin damping timescale so that
$
\tau = 100\Delta t.
$
Finally, we choose a value of~$\mu$ close to the mass of a single water layer,
$
\mu = 100\,\text{amu}/\text{nm}^2.
$

This interface dynamics is entirely fictitious.  However, it correctly samples configurations of the interface Boltzmann-weighted by the Hamiltonian~$H[\{h_\va\}]$.  This is true regardless of the exact values of $\mu$, $\Delta t$~and~$\tau$, so our choices have no effect on the results in the main text.  We have simply chosen reasonable values that do not lead to large discretization errors when solving the system's equations of motion.


\subsection{Surface-interface interactions}

The liquid-vapor-like interface interacts with the model surface via a potential~$H'_{\text{surf}}[\{h_\va\}]$.  In the atomistic simulations, the SAM sets up an interaction potential $U(\vr)$ felt by the atoms in the water molecules.  Below, we use the notation $\vr$ and $(x,y,z)$ interchangeably.  To model this interaction potential, we smear out the atomistic detail of the SAM and replace it with three elements:
\begin{itemize}
\item A uniform area density~$\muHead$ of Lennard-Jones sites (with length and energy scales $\sigmaHead$~and~$\epsHead$) in the $z = 0$ plane to represent the SAM head groups.
\item A uniform volume density~$\rhoTail$ of Lennard-Jones sites (with length and energy scales $\sigmaTail$~and~$\epsTail$) in the half-space $z < -\zeta$ to represent the SAM tail groups.
\item Coarse-graining the head-group atoms into a uniform area density results in a softer repulsive potential allowing the interface to penetrate far deeper into the model surface than would be possible in the actual SAM. To rectify this, we apply a strongly repulsive linear potential in the half-space $z < R_0$, where $R_0$ is the radius of the head group's hard core.  The repulsive potential is chosen to be $1\,\kB T$ when $1\,$nm$^2$ of interface penetrates the region $z < R_0$ by a ``skin depth''~$\delta$. 
\end{itemize}
The head groups are thus modeled by the following potential acting on a water molecule at position~$\vr$:
$$
U_{\text{head}}(x, y, z \geq R_0)
= \muHead \int_{-\infty}^\infty \text{d}x'\, \int_{-\infty}^\infty \text{d}y'\,   u_{\text{LJ}}\left(|\vr - \vr'|;\epsHead,\sigmaHead\right)\big|_{z' = 0},
$$
where
$
u_{\text{LJ}}(r;\epsilon,\sigma)=4\epsilon[(\sigma/r)^{12} - (\sigma/r)^6]
$
is the Lennard-Jones pair potential.
Similarly, the effect of the tail groups is captured by
$$
U_{\text{tail}}(x, y, z \geq R_0)
= \rhoTail \int_{-\infty}^\infty \text{d}x'\, \int_{-\infty}^\infty \text{d}y'\, \int_{-\infty}^{-\zeta} \text{d}z'\,  u_{\text{LJ}}\left(|\vr - \vr'|;\epsTail,\sigmaTail\right).
$$
Finally, the repulsive wall is modeled by the potential
$$
U_{\text{wall}}(x, y, z < R_0) = 
2 \rho_\ell^{-1} \cdot (1\,\kB T / 1\,\text{nm}^2) (R_0 - z) / \delta,
$$
where $\rho_\ell = 0.03333\,\text{\AA}^{-3}$ is the number density of liquid water.

These smeared interaction potentials depend only on~$z$, not on $x$~or~$y$.  As described in the main text, we also scale the head-group interaction by a parameter~$\eta$.  Putting everything together, we obtain an explicit expression for the surface-interface interaction potential,
$$
H'_{\text{surf}}[\{h_\va\}] = \rho_\ell\Delta^2 \sum_\va h'_{\text{surf}}(h_\va),
$$
where
\begin{align*}
h'_{\text{surf}}(h_\va) &= \int_{h_\va}^\infty\text{d}z\, \eta U_{\text{head}}(z) + U_{\text{tail}}(z) + U_{\text{wall}}(z),\\
&=\begin{cases}
\eta\pi\epsHead\muHead\sigmaHead^3 \left[ \frac 4 {45} (\sigmaHead/z)^{9}
- \frac 2 3 (\sigmaHead/z)^{3} \right]&\\
\quad + \pi\epsTail\rhoTail\sigmaTail^4 \left[ \frac 1 {90} (\sigmaHead/[z+\zeta])^{8}
- \frac{1}{3} (\sigmaHead/[z+\zeta])^{2} \right],&z \geq R_0,\\
h'_{\text{surf}}(R_0) + \rho_\ell^{-1} \cdot (1\,\kB T / 1\,\text{nm}^2) ([R_0 - z] / \delta)^2,&z < R_0.
\end{cases}
\end{align*}

To model the -CH$_3$ SAM in this paper, we chose the following values for the parameters
\begin{itemize}
\item The head groups are modeled as OPLS united-atom CH$_3$ groups interacting with SPC/E water, so $\sigmaHead = 3.5355\,$\AA\ and $\epsTail = 0.68976\,$kJ/mol.
\item The tail groups are modeled as OPLS united-atom CH$_2$ groups (sp$^3$-hybridized) interacting with SPC/E water, so $\sigmaTail = 3.5355\,$\AA\ and $\epsTail = 0.5664\,$kJ/mol.
\item The tail region is inset from the plane of the head groups by a distance equal to a CH$_2$-CH$_3$ bond length ($1.50\,$\AA), minus the van der Waals radius of a CH$_2$ group ($1.9525\,$\AA), so $\zeta = -0.4525\,$\AA.
\item The head group density is known from the atomistic SAM geometry to be $\muHead = 0.0462\,$\AA$^{-2}$.  The mass density of the SAM tails was estimated to be $935\,$kg/m$^3$~\cite{garde09pnas}, resulting in a CH$_2$ group number density of $\rhoTail = 0.0402\,$\AA$^{-3}$.
\item The equivalent hard sphere radius~$R_0$ of a -CH$_3$ group at room temperature was estimated to be $3.37\,$\AA~\cite{ajp_jpcb2010}.  It has a small temperature dependence, which we neglect.
\item The wall skin depth~$\delta$ was set to $0.1\,$\AA, which is small enough so that the repulsive potential is essentially a hard wall at~$z = R_0$, but large enough that we can propagate the interfacial dynamics with a reasonable timestep.
\end{itemize}

\subsection{Umbrella sampling}

Calculating $\mu^{\text{ex}}(V)$ from Equation~(2) of the main text as a thermal average $\langle P_v(0) \rangle$ over Boltzmann-weighted configurations of $h(\va)$ is impractical for large~$V$.  The configurations that dominate this average simply have a vanishingly small Boltzmann weight.  To solve this problem, and in analogy to what we do in atomistic simulations, we perform umbrella sampling on the size of the sub-volume~$v$ of the probe cavity~$V$ that is above the interface.

We begin by defining the volume $V$ corresponding to a probe cavity of dimensions $L\times L\times W$ as the set of points satisfying $|x|,|y| \leq L/2$~and~$R_0 \leq z \leq R_0+W$.  We then define~$v[\{h_\va\}]$ as the size of the sub-volume of~$V$ that is above the interface.  Using umbrella sampling and the multistate Bennet acceptance ratio method (MBAR)~\cite{ShirtsChodera2008}, we calculate the probability distribution for~$v$, $P(v)$, down to $v = 0$.  To do this, we use quadratic umbrellas defined by a center~$\bar v$ and width~$(\delta v)^2$, which result in the addition to the Hamiltonian of
$$
H'_{\text{umb}}[\{h_\va\}] = \kB T\, \frac{(v[\{h_\va\}] - \bar v)^2}{2 (\delta v)^2}.
$$
During each umbrella run, we also record the configurations~$\{h_\va\}$ which yield each observed value of~$v$.  We then approximate the right-hand side of Equation~(2) in the main text by summing over these configurations with appropriate weights, and obtain
$$
\mu^{\text{ex}}(V) \approx -\kB T \ln \frac{\sum_{\{h_\va\}} P_v(0) P(v[\{h_\va\}])}{\sum_{\{h_\va\}} P(v[\{h_\va\}])},
$$
where, as in the main text, the term~$P_v(0)$ depends on the interface configuration~$\{h_\va\}$, and the sum $\{h_\va\}$ is over all interface configurations in all the different umbrellas.  To evaluate $P_v(0)$, we implement discrete versions of the integrals defining $\langle N \rangle_v$~and~$\sigma_v$ as was done in Ref.~\cite{LLCW}.


\begin{thebibliography}{52}%
\makeatletter
\providecommand \@ifxundefined [1]{%
 \@ifx{#1\undefined}
}%
\providecommand \@ifnum [1]{%
 \ifnum #1\expandafter \@firstoftwo
 \else \expandafter \@secondoftwo
 \fi
}%
\providecommand \@ifx [1]{%
 \ifx #1\expandafter \@firstoftwo
 \else \expandafter \@secondoftwo
 \fi
}%
\providecommand \natexlab [1]{#1}%
\providecommand \enquote  [1]{``#1''}%
\providecommand \bibnamefont  [1]{#1}%
\providecommand \bibfnamefont [1]{#1}%
\providecommand \citenamefont [1]{#1}%
\providecommand \href@noop [0]{\@secondoftwo}%
\providecommand \href [0]{\begingroup \@sanitize@url \@href}%
\providecommand \@href[1]{\@@startlink{#1}\@@href}%
\providecommand \@@href[1]{\endgroup#1\@@endlink}%
\providecommand \@sanitize@url [0]{\catcode `\\12\catcode `\$12\catcode
  `\&12\catcode `\#12\catcode `\^12\catcode `\_12\catcode `\%12\relax}%
\providecommand \@@startlink[1]{}%
\providecommand \@@endlink[0]{}%
\providecommand \url  [0]{\begingroup\@sanitize@url \@url }%
\providecommand \@url [1]{\endgroup\@href {#1}{\urlprefix }}%
\providecommand \urlprefix  [0]{URL }%
\providecommand \Eprint [0]{\href }%
\@ifxundefined \urlstyle {%
  \providecommand \doi  [0]{\begingroup \@sanitize@url \@doi}%
  \providecommand \@doi [1]{\endgroup \@@startlink {\doibase
  #1}doi:\discretionary {}{}{}#1\@@endlink }%
}{%
  \providecommand \doi  [0]{doi:\discretionary{}{}{}\begingroup
  \urlstyle{rm}\Url }%
}%
\providecommand \doibase [0]{http://dx.doi.org/}%
\providecommand \Doi [0]{\begingroup \@sanitize@url \@Doi }%
\providecommand \@Doi  [1]{\endgroup\@@startlink{\doibase#1}\@@Doi}%
\providecommand \@@Doi [1]{#1\@@endlink}%
\providecommand \selectlanguage [0]{\@gobble}%
\providecommand \bibinfo  [0]{\@secondoftwo}%
\providecommand \bibfield  [0]{\@secondoftwo}%
\providecommand \translation [1]{[#1]}%
\providecommand \BibitemOpen [0]{}%
\providecommand \bibitemStop [0]{}%
\providecommand \bibitemNoStop [0]{.\EOS\space}%
\providecommand \EOS [0]{\spacefactor3000\relax}%
\providecommand \BibitemShut  [1]{\csname bibitem#1\endcsname}%
\bibitem [{\citenamefont {Tanford}(1973)}]{tanford_book}%
  \BibitemOpen
  \bibfield  {author} {\bibinfo {author} {\bibfnamefont {C.}~\bibnamefont
  {Tanford}},\ }\href@noop {} {\emph {\bibinfo {title} {The Hydrophobic Effect
  - Formation of Micelles and Biological Membranes}}}\ (\bibinfo  {publisher}
  {Wiley Interscience, New York},\ \bibinfo {year} {1973})\BibitemShut
  {NoStop}%
\bibitem [{\citenamefont {Kauzmann}(1959)}]{kauzmann}%
  \BibitemOpen
  \bibfield  {author} {\bibinfo {author} {\bibfnamefont {W.}~\bibnamefont
  {Kauzmann}},\ }\href@noop {} {\bibfield  {journal} {\bibinfo  {journal} {Adv.
  Prot. Chem.},\ }\textbf {\bibinfo {volume} {14}},\ \bibinfo {pages} {1}
  (\bibinfo {year} {1959})}\BibitemShut {NoStop}%
\bibitem [{\citenamefont {Stillinger}(1973)}]{stillinger}%
  \BibitemOpen
  \bibfield  {author} {\bibinfo {author} {\bibfnamefont {F.~H.}\ \bibnamefont
  {Stillinger}},\ }\href@noop {} {\bibfield  {journal} {\bibinfo  {journal} {J.
  Solution Chem.},\ }\textbf {\bibinfo {volume} {2}},\ \bibinfo {pages} {141}
  (\bibinfo {year} {1973})}\BibitemShut {NoStop}%
\bibitem [{\citenamefont {Lum}\ \emph {et~al.}(1999)\citenamefont {Lum},
  \citenamefont {Chandler},\ and\ \citenamefont {Weeks}}]{LCW}%
  \BibitemOpen
  \bibfield  {author} {\bibinfo {author} {\bibfnamefont {K.}~\bibnamefont
  {Lum}}, \bibinfo {author} {\bibfnamefont {D.}~\bibnamefont {Chandler}}, \
  and\ \bibinfo {author} {\bibfnamefont {J.~D.}\ \bibnamefont {Weeks}},\ }\Doi
  {10.1021/jp984327m} {\bibfield  {journal} {\bibinfo  {journal} {J. Phys.
  Chem. B},\ }\textbf {\bibinfo {volume} {103}},\ \bibinfo {pages} {4570}
  (\bibinfo {year} {1999})}\BibitemShut {NoStop}%
\bibitem [{\citenamefont {Chandler}(2005)}]{DC_nature05}%
  \BibitemOpen
  \bibfield  {author} {\bibinfo {author} {\bibfnamefont {D.}~\bibnamefont
  {Chandler}},\ }\href@noop {} {\bibfield  {journal} {\bibinfo  {journal}
  {Nature},\ }\textbf {\bibinfo {volume} {437}},\ \bibinfo {pages} {640}
  (\bibinfo {year} {2005})}\BibitemShut {NoStop}%
\bibitem [{\citenamefont {Mittal}\ and\ \citenamefont
  {Hummer}(2008)}]{mittal_pnas08}%
  \BibitemOpen
  \bibfield  {author} {\bibinfo {author} {\bibfnamefont {J.}~\bibnamefont
  {Mittal}}\ and\ \bibinfo {author} {\bibfnamefont {G.}~\bibnamefont
  {Hummer}},\ }\Doi {10.1073/pnas.0809029105} {\bibfield  {journal} {\bibinfo
  {journal} {P. Natl. Acad. Sci. U.S.A.},\ }\textbf {\bibinfo {volume} {105}},\
  \bibinfo {pages} {20130} (\bibinfo {year} {2008})}\BibitemShut {NoStop}%
\bibitem [{\citenamefont {Godawat}\ \emph {et~al.}(2009)\citenamefont
  {Godawat}, \citenamefont {Jamadagni},\ and\ \citenamefont
  {Garde}}]{garde09pnas}%
  \BibitemOpen
  \bibfield  {author} {\bibinfo {author} {\bibfnamefont {R.}~\bibnamefont
  {Godawat}}, \bibinfo {author} {\bibfnamefont {S.~N.}\ \bibnamefont
  {Jamadagni}}, \ and\ \bibinfo {author} {\bibfnamefont {S.}~\bibnamefont
  {Garde}},\ }\href@noop {} {\bibfield  {journal} {\bibinfo  {journal} {P.
  Natl. Acad. Sci. U.S.A.},\ }\textbf {\bibinfo {volume} {106}},\ \bibinfo
  {pages} {15119} (\bibinfo {year} {2009})}\BibitemShut {NoStop}%
\bibitem [{\citenamefont {Patel}\ \emph {et~al.}(2010)\citenamefont {Patel},
  \citenamefont {Varilly},\ and\ \citenamefont {Chandler}}]{ajp_jpcb2010}%
  \BibitemOpen
  \bibfield  {author} {\bibinfo {author} {\bibfnamefont {A.~J.}\ \bibnamefont
  {Patel}}, \bibinfo {author} {\bibfnamefont {P.}~\bibnamefont {Varilly}}, \
  and\ \bibinfo {author} {\bibfnamefont {D.}~\bibnamefont {Chandler}},\
  }\href@noop {} {\bibfield  {journal} {\bibinfo  {journal} {J. Phys. Chem.
  B},\ }\textbf {\bibinfo {volume} {114}},\ \bibinfo {pages} {1632} (\bibinfo
  {year} {2010})}\BibitemShut {NoStop}%
\bibitem [{\citenamefont {Shenogina}\ \emph {et~al.}(2009)\citenamefont
  {Shenogina}, \citenamefont {Godawat}, \citenamefont {Keblinski},\ and\
  \citenamefont {Garde}}]{garde09prl_HT}%
  \BibitemOpen
  \bibfield  {author} {\bibinfo {author} {\bibfnamefont {N.}~\bibnamefont
  {Shenogina}}, \bibinfo {author} {\bibfnamefont {R.}~\bibnamefont {Godawat}},
  \bibinfo {author} {\bibfnamefont {P.}~\bibnamefont {Keblinski}}, \ and\
  \bibinfo {author} {\bibfnamefont {S.}~\bibnamefont {Garde}},\ }\href@noop {}
  {\bibfield  {journal} {\bibinfo  {journal} {Phys. Rev. Lett.},\ }\textbf
  {\bibinfo {volume} {102}},\ \bibinfo {pages} {156101} (\bibinfo {year}
  {2009})}\BibitemShut {NoStop}%
\bibitem [{\citenamefont {Hummer}\ \emph {et~al.}(1996)\citenamefont {Hummer},
  \citenamefont {Garde}, \citenamefont {Garcia}, \citenamefont {Pohorille},\
  and\ \citenamefont {Pratt}}]{information_theory}%
  \BibitemOpen
  \bibfield  {author} {\bibinfo {author} {\bibfnamefont {G.}~\bibnamefont
  {Hummer}}, \bibinfo {author} {\bibfnamefont {S.}~\bibnamefont {Garde}},
  \bibinfo {author} {\bibfnamefont {A.~E.}\ \bibnamefont {Garcia}}, \bibinfo
  {author} {\bibfnamefont {A.}~\bibnamefont {Pohorille}}, \ and\ \bibinfo
  {author} {\bibfnamefont {L.~R.}\ \bibnamefont {Pratt}},\ }\href@noop {}
  {\bibfield  {journal} {\bibinfo  {journal} {P. Natl. Acad. Sci. U.S.A.},\
  }\textbf {\bibinfo {volume} {93}},\ \bibinfo {pages} {8951} (\bibinfo {year}
  {1996})}\BibitemShut {NoStop}%
\bibitem [{\citenamefont {Crooks}\ and\ \citenamefont
  {Chandler}(1997)}]{CrooksChandler1997}%
  \BibitemOpen
  \bibfield  {author} {\bibinfo {author} {\bibfnamefont {G.~E.}\ \bibnamefont
  {Crooks}}\ and\ \bibinfo {author} {\bibfnamefont {D.}~\bibnamefont
  {Chandler}},\ }\href@noop {} {\bibfield  {journal} {\bibinfo  {journal}
  {Phys. Rev. E},\ }\textbf {\bibinfo {volume} {56}},\ \bibinfo {pages} {4217}
  (\bibinfo {year} {1997})}\BibitemShut {NoStop}%
\bibitem [{\citenamefont {Rajamani}\ \emph {et~al.}(2005)\citenamefont
  {Rajamani}, \citenamefont {Truskett},\ and\ \citenamefont {Garde}}]{garde05}%
  \BibitemOpen
  \bibfield  {author} {\bibinfo {author} {\bibfnamefont {S.}~\bibnamefont
  {Rajamani}}, \bibinfo {author} {\bibfnamefont {T.~M.}\ \bibnamefont
  {Truskett}}, \ and\ \bibinfo {author} {\bibfnamefont {S.}~\bibnamefont
  {Garde}},\ }\href@noop {} {\bibfield  {journal} {\bibinfo  {journal} {P.
  Natl. Acad. Sci. U.S.A.},\ }\textbf {\bibinfo {volume} {102}},\ \bibinfo
  {pages} {9475} (\bibinfo {year} {2005})}\BibitemShut {NoStop}%
\bibitem [{\citenamefont {Acharya}\ \emph {et~al.}(2010)\citenamefont
  {Acharya}, \citenamefont {Vembanur}, \citenamefont {Jamadagni},\ and\
  \citenamefont {Garde}}]{acharya:faraday}%
  \BibitemOpen
  \bibfield  {author} {\bibinfo {author} {\bibfnamefont {H.}~\bibnamefont
  {Acharya}}, \bibinfo {author} {\bibfnamefont {S.}~\bibnamefont {Vembanur}},
  \bibinfo {author} {\bibfnamefont {S.~N.}\ \bibnamefont {Jamadagni}}, \ and\
  \bibinfo {author} {\bibfnamefont {S.}~\bibnamefont {Garde}},\ }\href@noop {}
  {\bibfield  {journal} {\bibinfo  {journal} {Faraday Discuss.},\ }\textbf
  {\bibinfo {volume} {146}},\ \bibinfo {pages} {353} (\bibinfo {year}
  {2010})}\BibitemShut {NoStop}%
\bibitem [{\citenamefont {Giovambattista}\ \emph {et~al.}(2008)\citenamefont
  {Giovambattista}, \citenamefont {Lopez}, \citenamefont {Rossky},\ and\
  \citenamefont {Debenedetti}}]{Giovambattista:PNAS:08}%
  \BibitemOpen
  \bibfield  {author} {\bibinfo {author} {\bibfnamefont {N.}~\bibnamefont
  {Giovambattista}}, \bibinfo {author} {\bibfnamefont {C.~F.}\ \bibnamefont
  {Lopez}}, \bibinfo {author} {\bibfnamefont {P.~J.}\ \bibnamefont {Rossky}}, \
  and\ \bibinfo {author} {\bibfnamefont {P.~G.}\ \bibnamefont {Debenedetti}},\
  }\href@noop {} {\bibfield  {journal} {\bibinfo  {journal} {P. Natl. Acad.
  Sci. U.S.A.},\ }\textbf {\bibinfo {volume} {105}},\ \bibinfo {pages} {2274}
  (\bibinfo {year} {2008})}\BibitemShut {NoStop}%
\bibitem [{\citenamefont {Garde}\ \emph {et~al.}(1996)\citenamefont {Garde},
  \citenamefont {Hummer}, \citenamefont {Garcia}, \citenamefont {Paulaitis},\
  and\ \citenamefont {Pratt}}]{garde96}%
  \BibitemOpen
  \bibfield  {author} {\bibinfo {author} {\bibfnamefont {S.}~\bibnamefont
  {Garde}}, \bibinfo {author} {\bibfnamefont {G.}~\bibnamefont {Hummer}},
  \bibinfo {author} {\bibfnamefont {A.~E.}\ \bibnamefont {Garcia}}, \bibinfo
  {author} {\bibfnamefont {M.~E.}\ \bibnamefont {Paulaitis}}, \ and\ \bibinfo
  {author} {\bibfnamefont {L.~R.}\ \bibnamefont {Pratt}},\ }\Doi
  {10.1103/PhysRevLett.77.4966} {\bibfield  {journal} {\bibinfo  {journal}
  {Phys. Rev. Lett.},\ }\textbf {\bibinfo {volume} {77}},\ \bibinfo {pages}
  {4966} (\bibinfo {year} {1996})}\BibitemShut {NoStop}%
\bibitem [{\citenamefont {Huang}\ and\ \citenamefont
  {Chandler}(2000)}]{HC_temp}%
  \BibitemOpen
  \bibfield  {author} {\bibinfo {author} {\bibfnamefont {D.~M.}\ \bibnamefont
  {Huang}}\ and\ \bibinfo {author} {\bibfnamefont {D.}~\bibnamefont
  {Chandler}},\ }\href@noop {} {\bibfield  {journal} {\bibinfo  {journal} {P.
  Natl. Acad. Sci. U.S.A.},\ }\textbf {\bibinfo {volume} {97}} (\bibinfo {year}
  {2000})}\BibitemShut {NoStop}%
\bibitem [{\citenamefont {Alejandre}\ \emph {et~al.}(1995)\citenamefont
  {Alejandre}, \citenamefont {Tildesley},\ and\ \citenamefont
  {Chapela}}]{water_gamma_vs_T}%
  \BibitemOpen
  \bibfield  {author} {\bibinfo {author} {\bibfnamefont {J.}~\bibnamefont
  {Alejandre}}, \bibinfo {author} {\bibfnamefont {D.~J.}\ \bibnamefont
  {Tildesley}}, \ and\ \bibinfo {author} {\bibfnamefont {G.~A.}\ \bibnamefont
  {Chapela}},\ }\href@noop {} {\bibfield  {journal} {\bibinfo  {journal} {J.
  Chem. Phys.},\ }\textbf {\bibinfo {volume} {102}},\ \bibinfo {pages} {4574}
  (\bibinfo {year} {1995})}\BibitemShut {NoStop}%
\bibitem [{\citenamefont {Athawale}\ \emph {et~al.}(2008)\citenamefont
  {Athawale}, \citenamefont {Sarupria},\ and\ \citenamefont
  {Garde}}]{garde08salt}%
  \BibitemOpen
  \bibfield  {author} {\bibinfo {author} {\bibfnamefont {M.~V.}\ \bibnamefont
  {Athawale}}, \bibinfo {author} {\bibfnamefont {S.}~\bibnamefont {Sarupria}},
  \ and\ \bibinfo {author} {\bibfnamefont {S.}~\bibnamefont {Garde}},\
  }\href@noop {} {\bibfield  {journal} {\bibinfo  {journal} {J. Phys. Chem.
  B},\ }\textbf {\bibinfo {volume} {112}},\ \bibinfo {pages} {5661} (\bibinfo
  {year} {2008})}\BibitemShut {NoStop}%
\bibitem [{\citenamefont {Maibaum}\ \emph {et~al.}(2004)\citenamefont
  {Maibaum}, \citenamefont {Dinner},\ and\ \citenamefont
  {Chandler}}]{MaibaumDinnerChandler2004}%
  \BibitemOpen
  \bibfield  {author} {\bibinfo {author} {\bibfnamefont {L.}~\bibnamefont
  {Maibaum}}, \bibinfo {author} {\bibfnamefont {A.~R.}\ \bibnamefont {Dinner}},
  \ and\ \bibinfo {author} {\bibfnamefont {D.}~\bibnamefont {Chandler}},\ }\Doi
  {10.1021/jp037487t} {\bibfield  {journal} {\bibinfo  {journal} {J. Phys.
  Chem. B},\ }\textbf {\bibinfo {volume} {108}},\ \bibinfo {pages} {6778}
  (\bibinfo {year} {2004})}\BibitemShut {NoStop}%
\bibitem [{\citenamefont {ten Wolde}\ and\ \citenamefont
  {Chandler}(2002)}]{tenWoldeChandler2002}%
  \BibitemOpen
  \bibfield  {author} {\bibinfo {author} {\bibfnamefont {P.~R.}\ \bibnamefont
  {ten Wolde}}\ and\ \bibinfo {author} {\bibfnamefont {D.}~\bibnamefont
  {Chandler}},\ }\href@noop {} {\bibfield  {journal} {\bibinfo  {journal} {P.
  Natl. Acad. Sci. U.S.A.},\ }\textbf {\bibinfo {volume} {99}},\ \bibinfo
  {pages} {6539} (\bibinfo {year} {2002})}\BibitemShut {NoStop}%
\bibitem [{\citenamefont {Miller}\ \emph {et~al.}(2007)\citenamefont {Miller},
  \citenamefont {Vanden-Eijnden},\ and\ \citenamefont {Chandler}}]{tommy}%
  \BibitemOpen
  \bibfield  {author} {\bibinfo {author} {\bibfnamefont {T.}~\bibnamefont
  {Miller}}, \bibinfo {author} {\bibfnamefont {E.}~\bibnamefont
  {Vanden-Eijnden}}, \ and\ \bibinfo {author} {\bibfnamefont {D.}~\bibnamefont
  {Chandler}},\ }\href@noop {} {\bibfield  {journal} {\bibinfo  {journal} {P.
  Natl. Acad. Sci. U.S.A.},\ }\textbf {\bibinfo {volume} {104}},\ \bibinfo
  {pages} {14559} (\bibinfo {year} {2007})}\BibitemShut {NoStop}%
\bibitem [{\citenamefont {Ferguson}\ \emph {et~al.}(2009)\citenamefont
  {Ferguson}, \citenamefont {Debenedetti},\ and\ \citenamefont
  {Panagiotopoulos}}]{pgd:jpcb:2009}%
  \BibitemOpen
  \bibfield  {author} {\bibinfo {author} {\bibfnamefont {A.~L.}\ \bibnamefont
  {Ferguson}}, \bibinfo {author} {\bibfnamefont {P.~G.}\ \bibnamefont
  {Debenedetti}}, \ and\ \bibinfo {author} {\bibfnamefont {A.~Z.}\ \bibnamefont
  {Panagiotopoulos}},\ }\href@noop {} {\bibfield  {journal} {\bibinfo
  {journal} {J. Phys. Chem. B},\ }\textbf {\bibinfo {volume} {113}},\ \bibinfo
  {pages} {6405} (\bibinfo {year} {2009})}\BibitemShut {NoStop}%
\bibitem [{\citenamefont {Huang}\ \emph {et~al.}(2005)\citenamefont {Huang},
  \citenamefont {Zhou},\ and\ \citenamefont {Berne}}]{HuangZhouBerne2005}%
  \BibitemOpen
  \bibfield  {author} {\bibinfo {author} {\bibfnamefont {X.}~\bibnamefont
  {Huang}}, \bibinfo {author} {\bibfnamefont {R.}~\bibnamefont {Zhou}}, \ and\
  \bibinfo {author} {\bibfnamefont {B.~J.}\ \bibnamefont {Berne}},\ }\href@noop
  {} {\bibfield  {journal} {\bibinfo  {journal} {J. Phys. Chem. B},\ }\textbf
  {\bibinfo {volume} {109}},\ \bibinfo {pages} {3546} (\bibinfo {year}
  {2005})}\BibitemShut {NoStop}%
\bibitem [{\citenamefont {Beverung}\ \emph {et~al.}(1999)\citenamefont
  {Beverung}, \citenamefont {Radke},\ and\ \citenamefont {Blanch}}]{Radke}%
  \BibitemOpen
  \bibfield  {author} {\bibinfo {author} {\bibfnamefont {C.~J.}\ \bibnamefont
  {Beverung}}, \bibinfo {author} {\bibfnamefont {C.~J.}\ \bibnamefont {Radke}},
  \ and\ \bibinfo {author} {\bibfnamefont {H.~W.}\ \bibnamefont {Blanch}},\
  }\href@noop {} {\bibfield  {journal} {\bibinfo  {journal} {Biophys. Chem.},\
  }\textbf {\bibinfo {volume} {81}},\ \bibinfo {pages} {59} (\bibinfo {year}
  {1999})}\BibitemShut {NoStop}%
\bibitem [{\citenamefont {Sethuraman}\ \emph {et~al.}(2004)\citenamefont
  {Sethuraman}, \citenamefont {Vedantham}, \citenamefont {Imoto}, \citenamefont
  {Przybycien},\ and\ \citenamefont {Belfort}}]{Belfort}%
  \BibitemOpen
  \bibfield  {author} {\bibinfo {author} {\bibfnamefont {A.}~\bibnamefont
  {Sethuraman}}, \bibinfo {author} {\bibfnamefont {G.}~\bibnamefont
  {Vedantham}}, \bibinfo {author} {\bibfnamefont {T.}~\bibnamefont {Imoto}},
  \bibinfo {author} {\bibfnamefont {T.}~\bibnamefont {Przybycien}}, \ and\
  \bibinfo {author} {\bibfnamefont {G.}~\bibnamefont {Belfort}},\ }\href@noop
  {} {\bibfield  {journal} {\bibinfo  {journal} {Proteins},\ }\textbf {\bibinfo
  {volume} {56}},\ \bibinfo {pages} {669} (\bibinfo {year} {2004})}\BibitemShut
  {NoStop}%
\bibitem [{\citenamefont {Nikolic}\ \emph {et~al.}(2011)\citenamefont
  {Nikolic}, \citenamefont {Baud}, \citenamefont {Rauscher},\ and\
  \citenamefont {Pomes}}]{pomes}%
  \BibitemOpen
  \bibfield  {author} {\bibinfo {author} {\bibfnamefont {A.}~\bibnamefont
  {Nikolic}}, \bibinfo {author} {\bibfnamefont {S.}~\bibnamefont {Baud}},
  \bibinfo {author} {\bibfnamefont {S.}~\bibnamefont {Rauscher}}, \ and\
  \bibinfo {author} {\bibfnamefont {R.}~\bibnamefont {Pomes}},\ }\href@noop {}
  {\bibfield  {journal} {\bibinfo  {journal} {Proteins},\ }\textbf {\bibinfo
  {volume} {79}},\ \bibinfo {pages} {1} (\bibinfo {year} {2011})}\BibitemShut
  {NoStop}%
\bibitem [{\citenamefont {Sharma}\ \emph {et~al.}(2010)\citenamefont {Sharma},
  \citenamefont {Berne},\ and\ \citenamefont {Kumar}}]{Berne:10:BJ}%
  \BibitemOpen
  \bibfield  {author} {\bibinfo {author} {\bibfnamefont {S.}~\bibnamefont
  {Sharma}}, \bibinfo {author} {\bibfnamefont {B.~J.}\ \bibnamefont {Berne}}, \
  and\ \bibinfo {author} {\bibfnamefont {S.~K.}\ \bibnamefont {Kumar}},\
  }\href@noop {} {\bibfield  {journal} {\bibinfo  {journal} {Biophys. J.},\
  }\textbf {\bibinfo {volume} {99}},\ \bibinfo {pages} {1157} (\bibinfo {year}
  {2010})}\BibitemShut {NoStop}%
\bibitem [{\citenamefont {Jamadagni}\ \emph {et~al.}(2009)\citenamefont
  {Jamadagni}, \citenamefont {Godawat}, \citenamefont {Dordick},\ and\
  \citenamefont {Garde}}]{SJ_jpcb09}%
  \BibitemOpen
  \bibfield  {author} {\bibinfo {author} {\bibfnamefont {S.~N.}\ \bibnamefont
  {Jamadagni}}, \bibinfo {author} {\bibfnamefont {R.}~\bibnamefont {Godawat}},
  \bibinfo {author} {\bibfnamefont {J.~S.}\ \bibnamefont {Dordick}}, \ and\
  \bibinfo {author} {\bibfnamefont {S.}~\bibnamefont {Garde}},\ }\href@noop {}
  {\bibfield  {journal} {\bibinfo  {journal} {J. Phys. Chem. B},\ }\textbf
  {\bibinfo {volume} {113}},\ \bibinfo {pages} {4093} (\bibinfo {year}
  {2009})}\BibitemShut {NoStop}%
\bibitem [{\citenamefont {Krone}\ \emph {et~al.}(2008)\citenamefont {Krone},
  \citenamefont {Hua}, \citenamefont {Soto}, \citenamefont {Zhou},
  \citenamefont {Berne},\ and\ \citenamefont {Shea}}]{shea08}%
  \BibitemOpen
  \bibfield  {author} {\bibinfo {author} {\bibfnamefont {M.~G.}\ \bibnamefont
  {Krone}}, \bibinfo {author} {\bibfnamefont {L.}~\bibnamefont {Hua}}, \bibinfo
  {author} {\bibfnamefont {P.}~\bibnamefont {Soto}}, \bibinfo {author}
  {\bibfnamefont {R.}~\bibnamefont {Zhou}}, \bibinfo {author} {\bibfnamefont
  {B.~J.}\ \bibnamefont {Berne}}, \ and\ \bibinfo {author} {\bibfnamefont
  {J.-E.}\ \bibnamefont {Shea}},\ }\href@noop {} {\bibfield  {journal}
  {\bibinfo  {journal} {J. Am. Chem. Soc.},\ }\textbf {\bibinfo {volume}
  {130}},\ \bibinfo {pages} {11066} (\bibinfo {year} {2008})}\BibitemShut
  {NoStop}%
\bibitem [{\citenamefont {Fenton}\ and\ \citenamefont
  {Horwich}(2003)}]{Chaperonins}%
  \BibitemOpen
  \bibfield  {author} {\bibinfo {author} {\bibfnamefont {W.}~\bibnamefont
  {Fenton}}\ and\ \bibinfo {author} {\bibfnamefont {A.}~\bibnamefont
  {Horwich}},\ }\href@noop {} {\bibfield  {journal} {\bibinfo  {journal} {Q.
  Rev. Biophys.},\ }\textbf {\bibinfo {volume} {36}},\ \bibinfo {pages} {229}
  (\bibinfo {year} {2003})}\BibitemShut {NoStop}%
\bibitem [{\citenamefont {England}\ \emph {et~al.}(2008)\citenamefont
  {England}, \citenamefont {Lucent},\ and\ \citenamefont {Pande}}]{Pande}%
  \BibitemOpen
  \bibfield  {author} {\bibinfo {author} {\bibfnamefont {J.}~\bibnamefont
  {England}}, \bibinfo {author} {\bibfnamefont {D.}~\bibnamefont {Lucent}}, \
  and\ \bibinfo {author} {\bibfnamefont {V.}~\bibnamefont {Pande}},\
  }\href@noop {} {\bibfield  {journal} {\bibinfo  {journal} {Curr. Opin. Struc.
  Biol.},\ }\textbf {\bibinfo {volume} {18}},\ \bibinfo {pages} {163} (\bibinfo
  {year} {2008})}\BibitemShut {NoStop}%
\bibitem [{\citenamefont {Jewett}\ and\ \citenamefont
  {Shea}(2010)}]{JEShea_rev}%
  \BibitemOpen
  \bibfield  {author} {\bibinfo {author} {\bibfnamefont {A.}~\bibnamefont
  {Jewett}}\ and\ \bibinfo {author} {\bibfnamefont {J.-E.}\ \bibnamefont
  {Shea}},\ }\href@noop {} {\bibfield  {journal} {\bibinfo  {journal} {Cell.
  Mol. Life Sci.},\ }\textbf {\bibinfo {volume} {67}},\ \bibinfo {pages} {255}
  (\bibinfo {year} {2010})}\BibitemShut {NoStop}%
\bibitem [{\citenamefont {Marchin}\ and\ \citenamefont
  {Berrie}(2003)}]{Marchin:Langmuir:2003}%
  \BibitemOpen
  \bibfield  {author} {\bibinfo {author} {\bibfnamefont {K.~L.}\ \bibnamefont
  {Marchin}}\ and\ \bibinfo {author} {\bibfnamefont {C.~L.}\ \bibnamefont
  {Berrie}},\ }\href@noop {} {\bibfield  {journal} {\bibinfo  {journal}
  {Langmuir},\ }\textbf {\bibinfo {volume} {19}},\ \bibinfo {pages} {9883}
  (\bibinfo {year} {2003})}\BibitemShut {NoStop}%
\bibitem [{\citenamefont {Anand}\ \emph {et~al.}(2011)\citenamefont {Anand},
  \citenamefont {Zhang}, \citenamefont {Linhardt},\ and\ \citenamefont
  {Belfort}}]{Belfort:Langmuir:2011}%
  \BibitemOpen
  \bibfield  {author} {\bibinfo {author} {\bibfnamefont {G.}~\bibnamefont
  {Anand}}, \bibinfo {author} {\bibfnamefont {F.}~\bibnamefont {Zhang}},
  \bibinfo {author} {\bibfnamefont {R.~J.}\ \bibnamefont {Linhardt}}, \ and\
  \bibinfo {author} {\bibfnamefont {G.}~\bibnamefont {Belfort}},\ }\href@noop
  {} {\bibfield  {journal} {\bibinfo  {journal} {Langmuir},\ }\textbf {\bibinfo
  {volume} {27}},\ \bibinfo {pages} {1830} (\bibinfo {year}
  {2011})}\BibitemShut {NoStop}%
\bibitem [{\citenamefont {Tian}\ \emph {et~al.}(2006)\citenamefont {Tian},
  \citenamefont {Cui}, \citenamefont {Schwarz}, \citenamefont {Estrada},\ and\
  \citenamefont {Kobayashi}}]{Tian:2006}%
  \BibitemOpen
  \bibfield  {author} {\bibinfo {author} {\bibfnamefont {F.}~\bibnamefont
  {Tian}}, \bibinfo {author} {\bibfnamefont {D.}~\bibnamefont {Cui}}, \bibinfo
  {author} {\bibfnamefont {H.}~\bibnamefont {Schwarz}}, \bibinfo {author}
  {\bibfnamefont {G.~G.}\ \bibnamefont {Estrada}}, \ and\ \bibinfo {author}
  {\bibfnamefont {H.}~\bibnamefont {Kobayashi}},\ }\href@noop {} {\bibfield
  {journal} {\bibinfo  {journal} {Toxicology in Vitro},\ }\textbf {\bibinfo
  {volume} {20}},\ \bibinfo {pages} {1202 } (\bibinfo {year}
  {2006})}\BibitemShut {NoStop}%
\bibitem [{\citenamefont {Makhatadze}\ and\ \citenamefont
  {Privalov}(1995)}]{privalov}%
  \BibitemOpen
  \bibfield  {author} {\bibinfo {author} {\bibfnamefont {G.}~\bibnamefont
  {Makhatadze}}\ and\ \bibinfo {author} {\bibfnamefont {P.}~\bibnamefont
  {Privalov}},\ }\href@noop {} {\bibfield  {journal} {\bibinfo  {journal} {Adv.
  Prot. Chem.},\ }\textbf {\bibinfo {volume} {47}},\ \bibinfo {pages} {307}
  (\bibinfo {year} {1995})}\BibitemShut {NoStop}%
\bibitem [{\citenamefont {Karajanagi}\ \emph {et~al.}(2004)\citenamefont
  {Karajanagi}, \citenamefont {Vertegel}, \citenamefont {Kane},\ and\
  \citenamefont {Dordick}}]{dordick}%
  \BibitemOpen
  \bibfield  {author} {\bibinfo {author} {\bibfnamefont {S.~S.}\ \bibnamefont
  {Karajanagi}}, \bibinfo {author} {\bibfnamefont {A.~A.}\ \bibnamefont
  {Vertegel}}, \bibinfo {author} {\bibfnamefont {R.~S.}\ \bibnamefont {Kane}},
  \ and\ \bibinfo {author} {\bibfnamefont {J.~S.}\ \bibnamefont {Dordick}},\
  }\href@noop {} {\bibfield  {journal} {\bibinfo  {journal} {Langmuir},\
  }\textbf {\bibinfo {volume} {20}},\ \bibinfo {pages} {11594} (\bibinfo {year}
  {2004})}\BibitemShut {NoStop}%
\bibitem [{\citenamefont {Patel}\ \emph {et~al.}(2011)\citenamefont {Patel},
  \citenamefont {Varilly}, \citenamefont {Chandler},\ and\ \citenamefont
  {Garde}}]{ajp_jstatphys}%
  \BibitemOpen
  \bibfield  {author} {\bibinfo {author} {\bibfnamefont {A.~J.}\ \bibnamefont
  {Patel}}, \bibinfo {author} {\bibfnamefont {P.}~\bibnamefont {Varilly}},
  \bibinfo {author} {\bibfnamefont {D.}~\bibnamefont {Chandler}}, \ and\
  \bibinfo {author} {\bibfnamefont {S.}~\bibnamefont {Garde}},\ }\href@noop {}
  {\bibfield  {journal} {\bibinfo  {journal} {J. Stat. Phys.},\ \bibinfo
  {pages} {submitted}} (\bibinfo {year} {2011})}\BibitemShut {NoStop}%
\bibitem [{\citenamefont {Berendsen}\ \emph {et~al.}(1987)\citenamefont
  {Berendsen}, \citenamefont {Grigera},\ and\ \citenamefont
  {Straatsma}}]{spce}%
  \BibitemOpen
  \bibfield  {author} {\bibinfo {author} {\bibfnamefont {H.~J.~C.}\
  \bibnamefont {Berendsen}}, \bibinfo {author} {\bibfnamefont {J.~R.}\
  \bibnamefont {Grigera}}, \ and\ \bibinfo {author} {\bibfnamefont {T.~P.}\
  \bibnamefont {Straatsma}},\ }\href@noop {} {\bibfield  {journal} {\bibinfo
  {journal} {J. Phys. Chem.},\ }\textbf {\bibinfo {volume} {91}},\ \bibinfo
  {pages} {6269} (\bibinfo {year} {1987})}\BibitemShut {NoStop}%
\bibitem [{\citenamefont {Essmann}\ \emph {et~al.}(1995)\citenamefont
  {Essmann}, \citenamefont {Perera}, \citenamefont {Berkowitz}, \citenamefont
  {Darden}, \citenamefont {Lee},\ and\ \citenamefont {Pedersen}}]{PME}%
  \BibitemOpen
  \bibfield  {author} {\bibinfo {author} {\bibfnamefont {U.}~\bibnamefont
  {Essmann}}, \bibinfo {author} {\bibfnamefont {L.}~\bibnamefont {Perera}},
  \bibinfo {author} {\bibfnamefont {M.~L.}\ \bibnamefont {Berkowitz}}, \bibinfo
  {author} {\bibfnamefont {T.}~\bibnamefont {Darden}}, \bibinfo {author}
  {\bibfnamefont {H.}~\bibnamefont {Lee}}, \ and\ \bibinfo {author}
  {\bibfnamefont {L.~G.}\ \bibnamefont {Pedersen}},\ }\href@noop {} {\bibfield
  {journal} {\bibinfo  {journal} {J. Chem. Phys.},\ }\textbf {\bibinfo {volume}
  {103}},\ \bibinfo {pages} {8577} (\bibinfo {year} {1995})}\BibitemShut
  {NoStop}%
\bibitem [{\citenamefont {Ryckaert}\ \emph {et~al.}(1977)\citenamefont
  {Ryckaert}, \citenamefont {Ciccotti},\ and\ \citenamefont
  {Berendsen}}]{SHAKE}%
  \BibitemOpen
  \bibfield  {author} {\bibinfo {author} {\bibfnamefont {J.-P.}\ \bibnamefont
  {Ryckaert}}, \bibinfo {author} {\bibfnamefont {G.}~\bibnamefont {Ciccotti}},
  \ and\ \bibinfo {author} {\bibfnamefont {H.~J.~C.}\ \bibnamefont
  {Berendsen}},\ }\href@noop {} {\bibfield  {journal} {\bibinfo  {journal} {J.
  Comp. Phys.},\ }\textbf {\bibinfo {volume} {23}},\ \bibinfo {pages} {327 }
  (\bibinfo {year} {1977})}\BibitemShut {NoStop}%
\bibitem [{\citenamefont {Widom}(1963)}]{Widom_jcp63}%
  \BibitemOpen
  \bibfield  {author} {\bibinfo {author} {\bibfnamefont {B.}~\bibnamefont
  {Widom}},\ }\href@noop {} {\bibfield  {journal} {\bibinfo  {journal} {J.
  Chem. Phys.},\ }\textbf {\bibinfo {volume} {39}},\ \bibinfo {pages} {2808 }
  (\bibinfo {year} {1963})}\BibitemShut {NoStop}%
\bibitem [{\citenamefont {Chandler}(1993)}]{gaussian_ft}%
  \BibitemOpen
  \bibfield  {author} {\bibinfo {author} {\bibfnamefont {D.}~\bibnamefont
  {Chandler}},\ }\href@noop {} {\bibfield  {journal} {\bibinfo  {journal}
  {Phys. Rev. E},\ }\textbf {\bibinfo {volume} {48}},\ \bibinfo {pages} {2898}
  (\bibinfo {year} {1993})}\BibitemShut {NoStop}%
\bibitem [{\citenamefont {Varilly}\ \emph {et~al.}(2011)\citenamefont
  {Varilly}, \citenamefont {Patel},\ and\ \citenamefont {Chandler}}]{LLCW}%
  \BibitemOpen
  \bibfield  {author} {\bibinfo {author} {\bibfnamefont {P.}~\bibnamefont
  {Varilly}}, \bibinfo {author} {\bibfnamefont {A.~J.}\ \bibnamefont {Patel}},
  \ and\ \bibinfo {author} {\bibfnamefont {D.}~\bibnamefont {Chandler}},\
  }\href@noop {} {\bibfield  {journal} {\bibinfo  {journal} {J. Chem. Phys.},\
  }\textbf {\bibinfo {volume} {134}},\ \bibinfo {pages} {074109} (\bibinfo
  {year} {2011})}\BibitemShut {NoStop}%
\bibitem [{\citenamefont {Narten}\ and\ \citenamefont
  {Levy}(1971)}]{NartenLevy1971}%
  \BibitemOpen
  \bibfield  {author} {\bibinfo {author} {\bibfnamefont {A.~H.}\ \bibnamefont
  {Narten}}\ and\ \bibinfo {author} {\bibfnamefont {H.~A.}\ \bibnamefont
  {Levy}},\ }\href@noop {} {\bibfield  {journal} {\bibinfo  {journal} {J. Chem.
  Phys.},\ }\textbf {\bibinfo {volume} {55}},\ \bibinfo {pages} {2263}
  (\bibinfo {year} {1971})}\BibitemShut {NoStop}%
\bibitem [{\citenamefont {Willard}\ and\ \citenamefont
  {Chandler}(2010)}]{WillardChandler2010}%
  \BibitemOpen
  \bibfield  {author} {\bibinfo {author} {\bibfnamefont {A.~P.}\ \bibnamefont
  {Willard}}\ and\ \bibinfo {author} {\bibfnamefont {D.}~\bibnamefont
  {Chandler}},\ }\href@noop {} {\bibfield  {journal} {\bibinfo  {journal} {J.
  Phys. Chem. B},\ }\textbf {\bibinfo {volume} {114}},\ \bibinfo {pages} {1954}
  (\bibinfo {year} {2010})}\BibitemShut {NoStop}%
\bibitem [{\citenamefont {Sedlmeier}\ \emph {et~al.}(2009)\citenamefont
  {Sedlmeier}, \citenamefont {Horinek},\ and\ \citenamefont
  {Netz}}]{SedlmeierEtAl2009}%
  \BibitemOpen
  \bibfield  {author} {\bibinfo {author} {\bibfnamefont {F.}~\bibnamefont
  {Sedlmeier}}, \bibinfo {author} {\bibfnamefont {D.}~\bibnamefont {Horinek}},
  \ and\ \bibinfo {author} {\bibfnamefont {R.~R.}\ \bibnamefont {Netz}},\
  }\href@noop {} {\bibfield  {journal} {\bibinfo  {journal} {Phys. Rev.
  Lett.},\ }\textbf {\bibinfo {volume} {103}},\ \bibinfo {pages} {136102}
  (\bibinfo {year} {2009})}\BibitemShut {NoStop}%
\bibitem [{\citenamefont {Vega}\ and\ \citenamefont
  {de~Miguel}(2007)}]{VegaMiguel2007}%
  \BibitemOpen
  \bibfield  {author} {\bibinfo {author} {\bibfnamefont {C.}~\bibnamefont
  {Vega}}\ and\ \bibinfo {author} {\bibfnamefont {E.}~\bibnamefont
  {de~Miguel}},\ }\href@noop {} {\bibfield  {journal} {\bibinfo  {journal} {J.
  Chem. Phys.},\ }\textbf {\bibinfo {volume} {126}},\ \bibinfo {pages} {154707}
  (\bibinfo {year} {2007})}\BibitemShut {NoStop}%
\bibitem [{\citenamefont {Jorgensen}\ \emph {et~al.}(1984)\citenamefont
  {Jorgensen}, \citenamefont {Madura},\ and\ \citenamefont
  {Swenson}}]{JorgensenMaduraSwenson1984}%
  \BibitemOpen
  \bibfield  {author} {\bibinfo {author} {\bibfnamefont {W.~L.}\ \bibnamefont
  {Jorgensen}}, \bibinfo {author} {\bibfnamefont {J.~D.}\ \bibnamefont
  {Madura}}, \ and\ \bibinfo {author} {\bibfnamefont {C.~J.}\ \bibnamefont
  {Swenson}},\ }\href@noop {} {\bibfield  {journal} {\bibinfo  {journal} {J.
  Am. Chem. Soc.},\ }\textbf {\bibinfo {volume} {106}},\ \bibinfo {pages}
  {6638} (\bibinfo {year} {1984})}\BibitemShut {NoStop}%
\bibitem [{\citenamefont {Buff}\ \emph {et~al.}(1965)\citenamefont {Buff},
  \citenamefont {Lovett},\ and\ \citenamefont
  {Stillinger}}]{BuffLovettStillinger1965}%
  \BibitemOpen
  \bibfield  {author} {\bibinfo {author} {\bibfnamefont {F.}~\bibnamefont
  {Buff}}, \bibinfo {author} {\bibfnamefont {R.}~\bibnamefont {Lovett}}, \ and\
  \bibinfo {author} {\bibfnamefont {F.~H.}\ \bibnamefont {Stillinger}},\
  }\href@noop {} {\bibfield  {journal} {\bibinfo  {journal} {Phys. Rev.
  Lett.},\ }\textbf {\bibinfo {volume} {15}},\ \bibinfo {pages} {621} (\bibinfo
  {year} {1965})}\BibitemShut {NoStop}%
\bibitem [{Note1()}]{Note1}%
  \BibitemOpen
  \bibinfo {note} {The constraint on the magnitude of ${\protect \mathbf {k}}$
  ensures that no Nyquist modes, i.e., modes with $k_x$~or~$k_y$ equal to $\pm
  \pi /D$, are ever excited. If they were included, these modes would also be
  purely real, and the variance of the real component of their noise terms
  would likewise be twice that of the real component of the interior
  modes.}\BibitemShut {Stop}%
\bibitem [{\citenamefont {Shirts}\ and\ \citenamefont
  {Chodera}(2008)}]{ShirtsChodera2008}%
  \BibitemOpen
  \bibfield  {author} {\bibinfo {author} {\bibfnamefont {M.~R.}\ \bibnamefont
  {Shirts}}\ and\ \bibinfo {author} {\bibfnamefont {J.~D.}\ \bibnamefont
  {Chodera}},\ }\href@noop {} {\bibfield  {journal} {\bibinfo  {journal} {J.
  Chem. Phys.},\ }\textbf {\bibinfo {volume} {129}},\ \bibinfo {pages} {124105}
  (\bibinfo {year} {2008})}\BibitemShut {NoStop}%
\end{thebibliography}
%

\end{document}